\documentclass{aa}  
\usepackage{graphicx}
\usepackage{longtable,lscape}
\usepackage{txfonts}
\begin{document}
\def\cgs{erg cm$^{-2}$ s$^{-1}$}
\def\gs{\hbox{\raise0.5ex\hbox{$>\lower1.06ex\hbox{$\kern-1.07em{\sim}$}$}}} % maggiore uguale circa

   \title{Black Hole Growth and Starburst Activity at z=0.6-4 in the Chandra Deep Field South}

   \subtitle{Host galaxies properties of obscured AGN}

   \author{M. Brusa          \inst{1}
          \and
          F. Fiore \inst{2} 
           \and 
          P. Santini \inst{2} 
            \and          
          A. Grazian \inst{2}
            \and
	  A. Comastri \inst{3}
	    \and
	  G. Zamorani \inst{3}
	    \and 
          G. Hasinger \inst{1,4} 
            \and 
          A. Merloni \inst{1,5} 
          \and
          F. Civano \inst{6} 
          \and 
          A. Fontana \inst{2}
          \and
          V. Mainieri \inst{7}
          }

   \offprints{marcella@mpe.mpg.de}

   \institute{Max-Planck-Institute for extraterrestrial Physik,
  Giessenbachstrasse, 1, D-85748, Garching bei M\"unchen, Germany \email{marcella@mpe.mpg.de}
         \and
             INAF --  Osservatorio Astronomico di Roma, via Frascati 33,
Monteporzio-Catone (Roma), I-00040, Italy \email{fiore@oa-roma.inaf.it}
        \and
             INAF --  Osservatorio Astronomico di Bologna, via Ranzani 1
40127 Bologna, Italy 
       \and 
Max-Planck-Institute for Plasma Physics, Boltzmannstrasse 2, D-85748, Garching, Germany
  \and
  Excellence Cluster Universe, 
  Boltzmannstrasse 2, D-85748, Garching bei Muenchen, Germany
    \and
      Harvard-Smithsonian Center for Astrophysics, 60 Garden
Street, Cambridge, MA 02138 
      \and
   European Southern Observatory, Karl-Schwarzschild-str. 2, 85748 Garching bei M\"unchen, Germany
}
 
   \date{Received 2 April 2009; accepted 22 September 2009}

% \abstract{}{}{}{}{} 
% 5 {} token are mandatory
 
  \abstract
 {} %leave it empty if necessary  
   {The co-evolution of host galaxies and the active black holes which 
    reside in their centre is one of the most important topics in modern 
    observational cosmology. Here we present a study of the properties 
   of obscured Active Galactic Nuclei (AGN) detected in the CDFS 1Ms 
   observation and their host galaxies.} 
   {We limited the analysis to the MUSIC area, for which deep K-band 
    observations obtained with ISAAC@VLT are available, ensuring accurate 
   identifications of the counterparts of the X--ray sources as well as 
   reliable determination of photometric redshifts and galaxy parameters, 
   such as stellar masses and star formation rates. 
   In particular, we: 1) refined the X-ray/infrared/optical association 
   of 179 sources in the MUSIC area detected in the Chandra observation; 
   2) studied the host galaxies observed and rest frame colors and properties.} 
  % resultsous heading (mandatory)
  {We found that  X--ray selected (L$_X\gs10^{42}$ erg s$^{-1}$)  
AGN show Spitzer colors consistent with both AGN and starburst dominated
infrared continuum; the latter would not have been selected as AGN
from infrared diagnostics. 
The host galaxies of X--ray selected obscured AGN are all massive
(M$_*>10^{10}$ M$\odot$) 
and, in 50\% of the cases, are also actively forming stars (1/SSFR$<$ t$_{Hubble}$) 
in dusty environments. 
The median L/LEdd value of the active 
nucleus is between 2\% and 10\% depending on the assumed 
M$_{BH}$/M$_{*}$ ratio. %, $\geq$ the local value. 
Finally, we found that the
X--ray selected AGN fraction increases with the stellar mass up to a value
of $\sim30$\% at z$>1$ and M$_*>3\times10^{11}$ M$\odot$, a fraction significantly
higher than in the local Universe for AGN of similar luminosities.}  
   {}


   \maketitle
%
%________________________________________________________________

\section{Introduction}

Galaxy interactions, and more in general the large scale structure
(LSS) galaxy environment, are thought to play a major role in
regulating both star-formation and accretion onto nuclear
super-massive black holes (SMBHs). 
This implies that the full understanding of galaxy evolution requires a
good knowledge of the SMBH census through cosmic time. 
In particular, feedback between the central SMBHs
in their active phases and the interstellar medium is likely
to affect strongly the evolution of their host galaxies.

    A short, powerful but highly obscured growth phase of
both SMBHs and their host galaxies is predicted by many models for the
co-evolution of galaxies and AGN (Silk \& Rees 1998, Fabian 1999,
Granato et al. 2004, Di Matteo et al. 2005, Menci et al. 2008). 
This phase ends when strong AGN winds and shocks heat
the interstellar medium, blowing away the dust and gas  
and inhibiting further star-formation in the AGN host
galaxies.  
According to this ``evolutionary sequence'' (e.g. Hopkins et al. 2008),
highly obscured AGN should be associated to young galaxies 
in the process of assembling most of their stellar mass 
through significant episodes of star formation.
On the contrary, unobscured AGN should be associated to galaxies with 
low or absent episodes of star-formation, given that most of the 
gas and dust responsible for the star formation has been blown away
by the effect of AGN feedback.
Many observational evidences (see Alexander 
et al. 2005, Page et al. 2004, Stevens et al. 2006) and theoretical
arguments (Menci et al. 2008 and references therein) in favor of 
the evolutionary sequence do exist. These results challenge our 20-years old
AGN view, in which the differences we see in different classes of
sources - especially between "obscured" and "unobscured" ones - 
are simply due to orientations effects (Antonucci et al. 1985, Antonucci 2003,
Urry \& Padovani 1995). \\
A correct and complete identification of obscured and highly obscured AGN 
at high redshift is therefore crucial 
because they may represent the first, still little explored, phase of 
the common growth of both SMBHs and their host galaxies. 

Moderately and highly obscured AGN at z$\sim2$, where most of the 
accretion and star-formation processes are on-going, 
offer an ideal tool for a direct test of
feedback models. In these objects the nuclear light is completely
blocked or strongly reduced and does not overshine the galaxy
optical and near-infrared light. 
This gives us the possibility to study host galaxy morphologies,
colors and spectral energy distributions without the difficulty of
disentangling star-light from nuclear light. We can therefore study
galaxy properties during active phases, e.g. {\it when nuclear
feedback should be in action}.

Obscured AGN at cosmological distances are however usually faint in
the observed optical bands, because the UV rest frame is strongly
reduced by dust extinction. 
On the other hand, moderately obscured AGN
(or Compton thin, $10^{22}<$N$_H<10^{24}$ cm$^{-2}$) are common in
X-ray images (Bauer et al. 2004, Comastri \& Brusa 2008 and reference
therein), making up $\sim$50\% of the full X-ray population at fluxes
$<10^{-14}$ \cgs (Gilli, Comastri \& Hasinger 2007) once the
contribution from the normal starforming galaxy population is removed 
(Ranalli et al. 2005).
Compton thick AGN (CT, N$_H\gs10^{24}$ cm$^{-2}$)
are faint also in the X-ray band, because photoelectric absorption and
Compton scattering strongly reduce the X-ray flux up to 
$>10$ keV, and over the entire X--ray range if N$_H\gs10^{25}$ cm$^{-2}$. 
Indeed, only a dozen CT AGN is present in the deepest images
of the X-ray sky, the Chandra Deep Fields (Norman et al. 2002, 
Tozzi et al.  2006, Georgantopoulos et al. 2008).

Identification of the correct counterparts of obscured AGN at
cosmological redshift is not trivial. These objects are faint in the
optical band, because 1) the intrinsic AGN emission is absorbed by the
surrounding material, and 2) the host galaxy light is strongly reduced
by cosmological dimming (L$^*$galaxies at z=1-2 have R$\sim24-25$).
The probability to find by chance a galaxy of these magnitudes in the
Chandra error boxes is not negligible.  The identification process is
made easier by using deep near infrared images, because the surface
density of these sources is smaller than that of optical ones,
bringing the probability of finding a galaxy by chance in the Chandra
error boxes to comfortably small values.  Moreover, the K-band flux is
more tightly correlated with the X-ray flux than the optical
(obscured) one.  For this reason, we have reanalyzed the
identifications of the Chandra Deep Field South (CDFS)
1 Ms sources (Alexander et al. 2003) in
the area covered by sensitive K band and IRAC 3.6$\mu$m and 4.5$\mu$m
observations (the GOODS MUSIC area, Grazian et al. 2006).

The excellent multiband photometry available in this area allows the
determination of a reliable photometric redshift for the faint sources
not reachable by optical spectroscopy.  Once identified the
counterparts of the X-ray sources and their redshift, it is possible to
proceed to a detailed study of the properties of their host
galaxies. We present here a study of the properties of the host
galaxies of X-ray obscured AGN, such as their morphology and close
environment, optical and infrared rest frame colors, mid infrared to
optical spectral energy distributions. Galaxy masses and
star-formation rates are derived from the fit of galaxy templates to
the optical to near infrared spectral energy distributions. We finally
compare the masses and star-formation rates of obscured AGN host
galaxies to those of inactive galaxies in the field selected in both
optical and infrared bands.

The paper is organized as follows: Sect. 2 presents the CDFS datasets,
the X-ray to optical/infrared association and the obscured AGN sample.
Section 3 presents the observed frame colors of the obscured AGN
sample.  Section 4 presents the host galaxies properties (masses and
star-formation rates, SFRs ) of the sample and our estimates of the
fraction of AGN in mass selected samples.  Section 5 presents a
a discussion of the results and Section 6 outlines a summary of 
the most important points.  A cosmology with $H_0=70$ km s$^{-1}$ Mpc$^{-1}$,
$\Omega_M$=0.3, $\Omega_{\Lambda}=0.7$ is adopted throughout
(Spergel et al. 2003).

\section{Obscured AGN selection in the CDFS}

\subsection{X-ray data: catalogs and multiband identifications}

%----------------------------------------------------------- S_vib
   \begin{figure}
   \includegraphics[width=8.5cm]{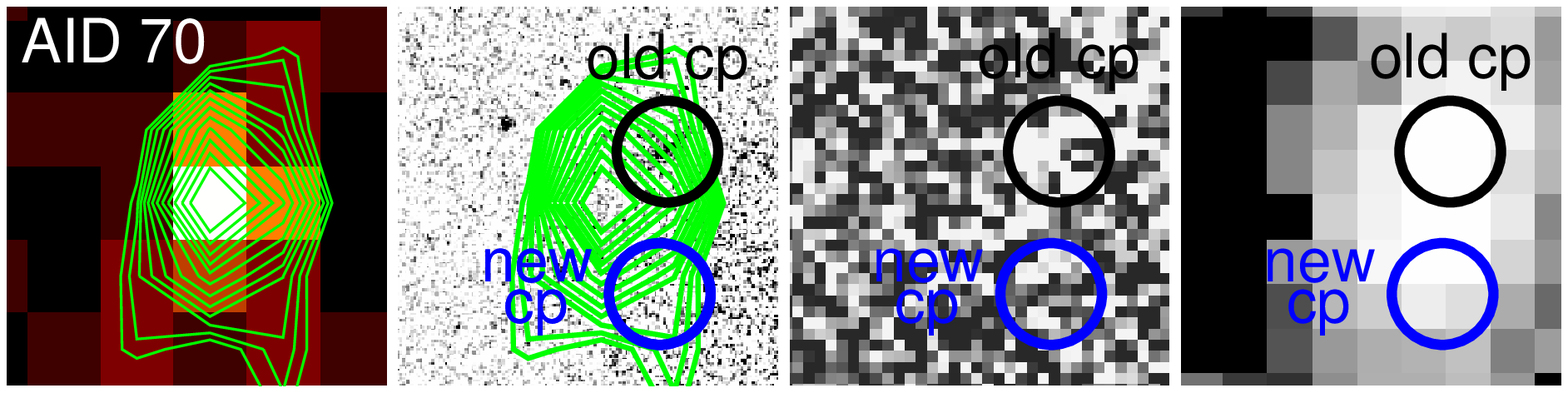}
   \includegraphics[width=8.5cm]{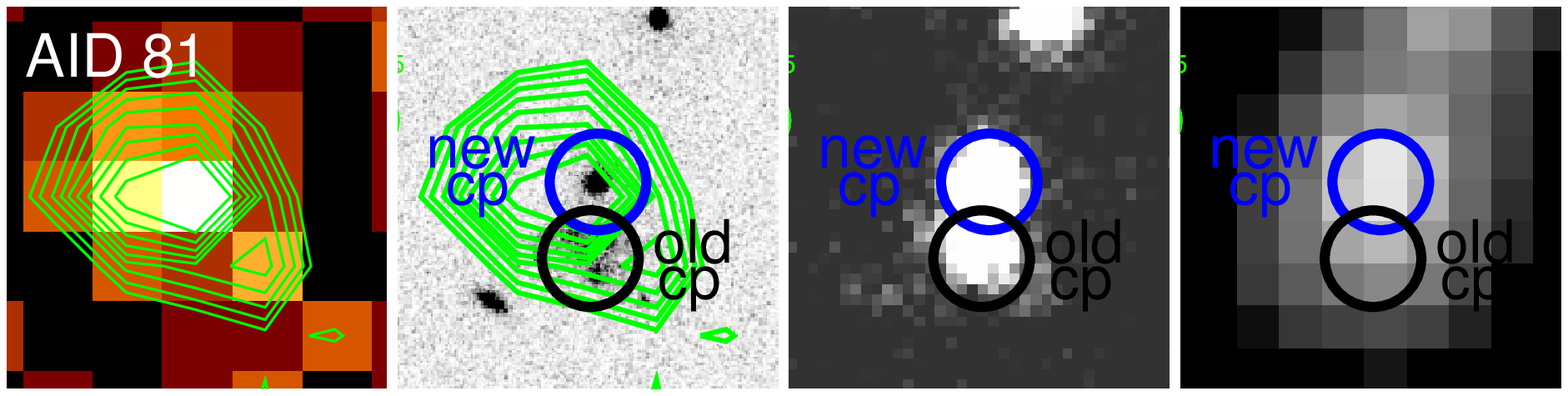}
   \includegraphics[width=8.5cm]{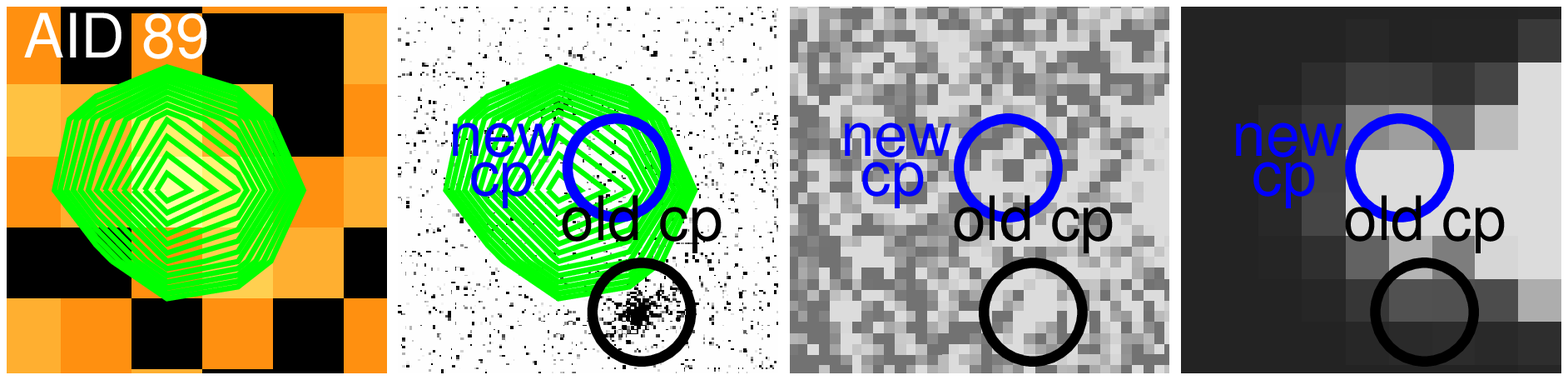}
   \includegraphics[width=8.5cm]{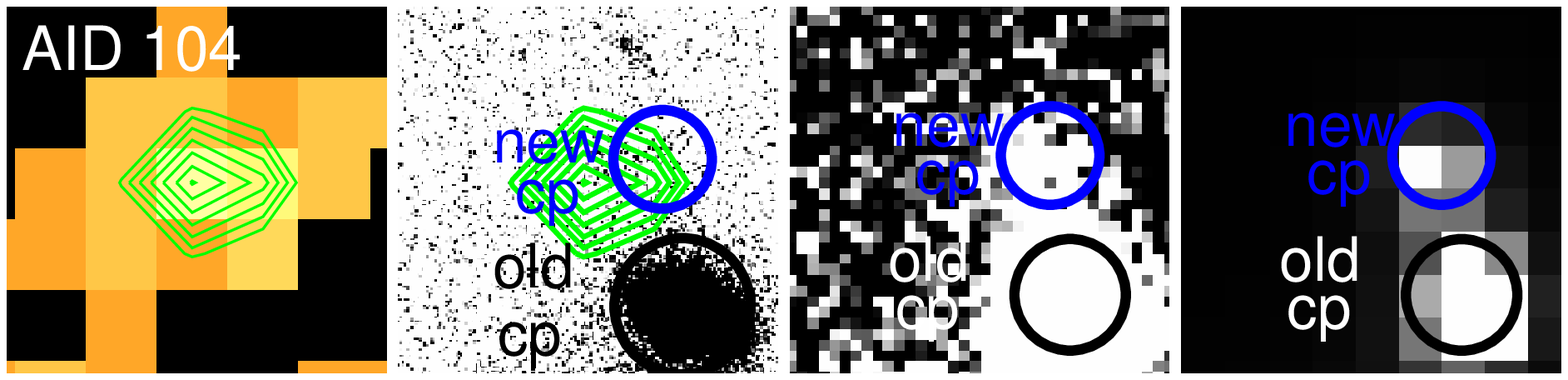}
   \includegraphics[width=8.5cm]{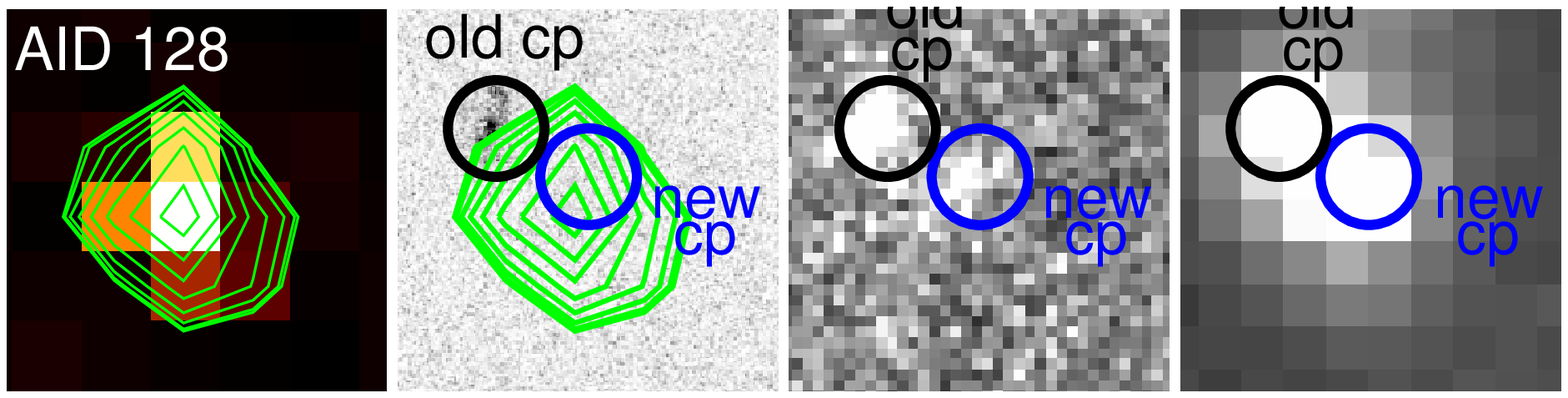}
   \includegraphics[width=8.5cm]{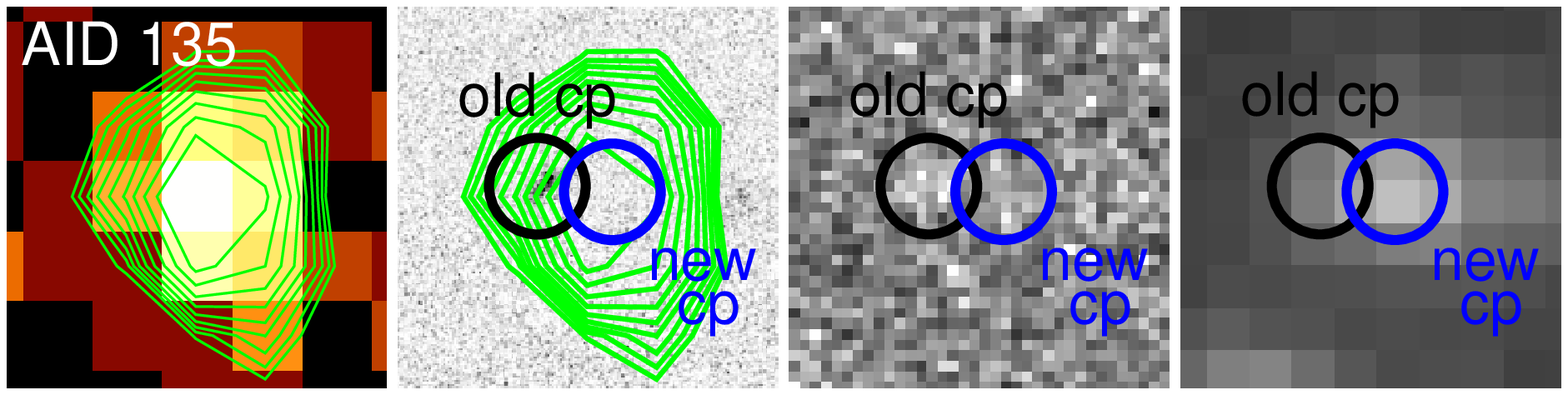}
   \includegraphics[width=8.5cm]{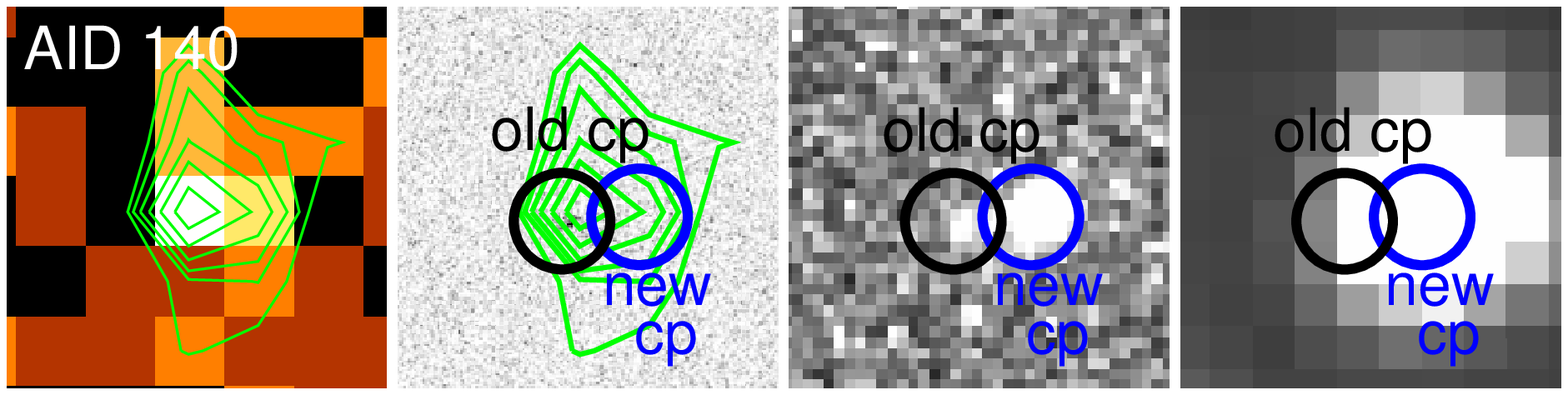}
   \includegraphics[width=8.5cm]{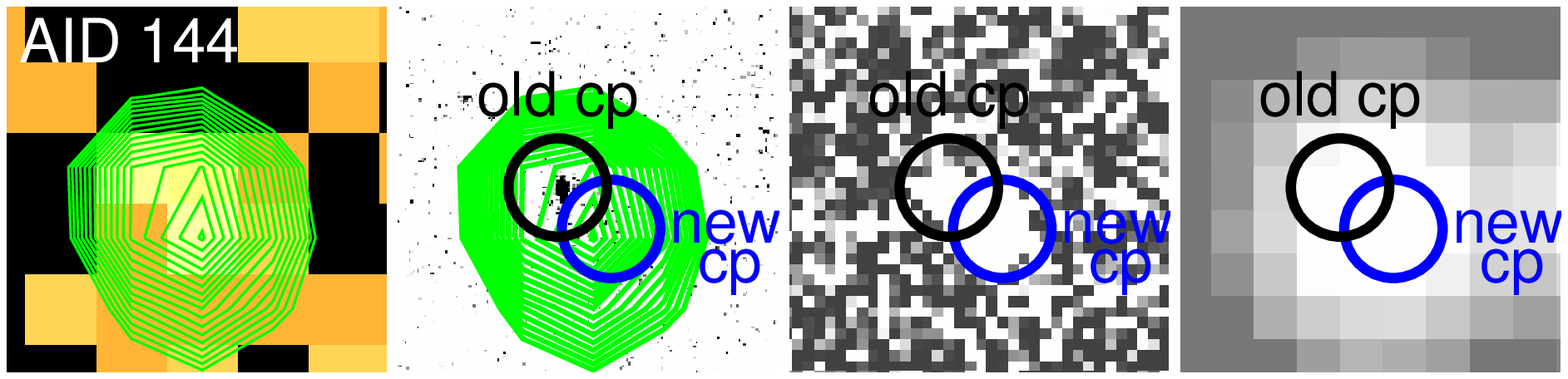}
   \includegraphics[width=8.5cm]{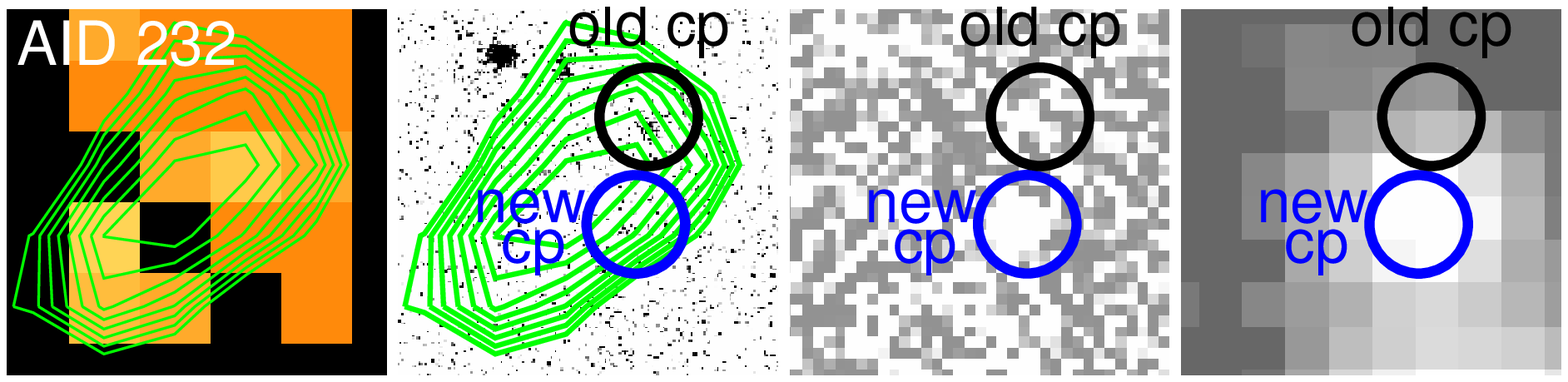}
      \caption{Chandra, ACS ($z-$band), K-band and IRAC-4.5 micron cutouts (5"x5") of the 9 sources identified with a counterpart different from that 
proposed in the literature. The blue circle marks the new counterpart, 
while the black circle marks the old counterpart. Smoothed countours
from the X-ray image are superimposed on the ACS image.}
         \label{FigVibStab}
   \end{figure}
%
%______________________________________________________________

X--ray data for a total exposure time of $\sim$1 Msec in the GOODS
field have been obtained by the CDFS consortium and are publicly
available.  The source catalog has been published by Giacconi et
al. (2002), including the basic X-ray properties in three different
X-ray bands (source counts, fluxes, exposure times). 
Subsequently, Alexander et al. (2003, hereinafter A03) published a re-analysis of
the CDFS data that led mainly to an improved astrometry of the X-ray
positions with some differences in the number of detected sources at
the very faint counts level (see A03 for more details). \\ Optical
identifications for the sources detected by Giacconi et al. (2002),
including the position of the candidate counterparts, optical
magnitudes and the spectroscopic redshifts from a dedicated VLT
campaign, are reported in Szokoly et al. (2004). Photometric redshifts
based on good quality optical photometry (VLT and ACS/HST) and deep
infrared (ISAAC/VLT) observations are available for all but four CDFS
sources without spectroscopic redshift (Zheng et al. 2004; Mainieri et
al. 2005), leading to a virtually complete catalog of identified
X--ray sources. Using this catalog and redshifts information, 
Tozzi et al. (2006) derived the rest-frame X--ray properties for
the 1 Msec CDFS sources. 
IRAC photometry, photometric redshifts and radio properties of different 
subsamples of the CDFS X-ray  sources have been published in 
others works appeared in the literature in the past few years, e.g. 
Rigby et al. (2005), Alonso-Herrero et al. (2006), Georgantopoulos et al. (2007), 
Rovilos \& Georgantopoulos (2007) and Tozzi et al.(2009). 

These published works rely on the association of the X-ray sources to a
candidate {\it optical} counterpart. However, given that the optical flux of
the X-ray counterparts can be strongly suppressed by dust absorption, an
identification process that uses as a prior a catalog generated at a redder
wavelength (K-band or IRAC bands) can be more efficient in isolating the 
correct association to the X-ray source (see, e.g. Brusa et al. 2007).  
The lower surface density of IRAC
sources with respect to faint optical objects is widely used 
as an argument to indicate that, in case
of ambiguous association between  two sources, the most likely candidate is
the one detected at the brightest IRAC fluxes (whatever the optical flux).\\

In order to quantify how (and if) the use of catalogs selected in different
bands leads to significantly different X-ray to optical associations, we
decided to perform again the identification of the Chandra sources, 
using a new version of the multi-color
GOODS-MUSIC sample (GOODS MUlticolor Southern
Infrared Catalog; Grazian et al. 2006, De Santis et al. 2007), 
extracted from the public data of the GOODS-South survey (Giavalisco 
et al. 2004), and updated as described in Santini et al. (2009).   
The 15-bands multi-wavelength coverage ranges from 0.35 to 24 $\mu$m,
as a result of the combination of images coming from different
instruments (2.2ESO, VIMOS/VLT, ACS/HST, ISAAC/VLT, IRAC/Spitzer and
MIPS/Spitzer). In the following we will refer to the catalog with full
coverage in all bands. Such a catalog covers an area of 143.2 arcmin$^2$
and contains 15208 sources. After culling Galactic stars, it
contains 14999, extragalactic objects selected in the z band, in the Ks band and
at 4.5 $\mu$m (see Santini et al. 2009 for further details). \\ A
total of 183 X-ray sources in the A03 catalog are found within the
GOODS-MUSIC area. Of these, 3 objects are associated with stars (ID
\#120,279,289 in A03) and an additional source is most likely a Ultra Lumimous 
X-ray source (ULX) in
a nearby galaxy (ID \#213). Therefore, the number of extragalactic
sources (AGN or galaxies) in the MUSIC area is 179, 154 in common with Giacconi et 
al. (2002) and an additional 25 detected only by A03\footnote{We
do not discuss here the 10 sources in the MUSIC area present only in
Giacconi et al. (2002)}. \\ 167 of the 179 AO3 sources in the MUSIC
area are in common with the catalog obtained from the full 2Msec
exposure (Luo et al. 2008).  We have visually inspected and carefully
checked all the 12 sources missing in Luo et al. (2008).  One of these
sources, \#225 is most likely a spurious detection on the wings of the
Chandra PSF of a brighter source (\#224). Other three sources,
\#183, \#215 and \#239, lie in problematic areas, close to a CCD gap and
are probably spurious detection in AO3. These four sources are flagged
as likely spurious in Table 1. The other 8 sources present in the AO3
catalog but not in the Luo et al. (2008) catalog are probably real
sources, which have not been detected in the full 2Msec exposure
because of variability or a more conservative detection threshold in
Luo et al. (2008), or both.

%----------------------------------------------------------- S_vib
   \begin{figure} \includegraphics[width=8cm]{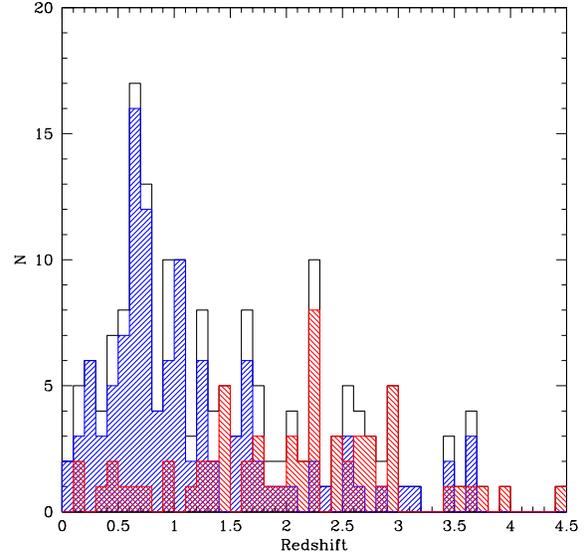} \caption{
Redshift distribution of the X--ray sources. Open
   histogram: all sources; right diagonal (blue) shaded histogram: 
   spectroscopic sample; left diagonal (red) shaded 
   histogram: photometric sample } \label{FigVibStab} \end{figure}
%
%______________________________________________________________

We proceeded for the association of the counterparts to the X-ray
sources as follows.
First, we used a statistical method (the
``likelihood ratio technique'', following the approach described by
Brusa et al. (2007) to isolate the most likely z-band, K-band and IRAC
counterparts from the z-, K- and 4.5 micron selected catalogs in the
MUSIC area. The method calculates the probability 
that a source is the correct association by weighing the information
on the X--ray to optical distance, the surface density of (possible)
false coincidence background objects and the brightness of the 
chosen counterpart. With this method we were able to isolate the
most obvious associations, and, at the same time to pick up the 
objects with 2 or more different counterparts with comparable likelihood in 
the same or different bands ($<10\%$ of the entire sample).
Then, we visually inspected all the proposed 179
associations in order to resolve possible ambiguous cases.  The
result of our work is summarized in Table 1. For each X-ray source we
report the X-ray identificator from A03 (AID, column 1), the X-ray
identificator from G02/S04 (XID, column 2), the MUSIC identificator
(OID, column 3), the positions of the proposed counterpart (column
4,5), the distance of the chosen counterpart from the X-ray centroid
(column 6), a flag for the identification (column 7), the observed
X-ray fluxes in the hard (2-10 keV) and soft (0.5-2 keV)
bands (column 8 and 9), the redshift of the source 
(column 10, see next section) and the rest frame 2-10 keV luminosity
(column 11).  
The X-ray fluxes (in erg cm$^{-2}$ s$^{-1}$) have been derived 
from the counts reported in A03, using the following conversion factors:

$$ F(2-10keV)=2.8\times10^{-11} Counts(2-8 keV)$$
$$ F(0.5-2keV)=5.6\times10^{-12} Counts(0.5-2keV$$

for a power law spectrum with energy index equal to 0.4,
i.e. the average value for deep fields sources (see, e.g., Tozzi 
et al. 2001)\footnote{We haven't used the fluxes tabulated
by A03 because they have been calculated with different 
(source-by-source) energy indeces and make it difficult to correct for 
the absorption and derive rest-frame, unobscured luminosity (see 2.2).}

Of the 154 published counterparts (first and second part
of the table, mostly from Szokoly et al. 2004, but see also 
notes in Table 1), we were able to confirm 144 X-ray to optical
associations (93.5\%, sources with flag = 0), we changed 9 X-ray to
optical associations (5.8\%, flag = 1)), while for the remaining case
(AID 186) we were not able to assign a reliable IRAC or
optical counterpart (0.6\%, i.e. neither to confirm the published
counterpart nor to propose a new counterpart).  The fraction of
previously misidentified sources rises up to 15\% when the optically faint
($z>24$) sample is considered (9/61). \\ 
Figure 1 shows the Chandra, ACS
($z-$band), K and IRAC-4.5 micron cutouts for the 9 sources we
identify with a counterpart different from that proposed in the
literature (see again notes in Table 1).
 The cyan circle marks the new counterpart, while the
yellow circle marks the old counterpart. In almost all the cases the
new proposed counterpart is absent in the optical images. \\ For the
remaining 25 sources detected only in A03, we were able to
unambiguously identify the counterpart for 23 of them.  The optical
counterparts of these 23 sources are reported in the third part of
table 1.

We computed the rest frame, absorption corrected 
2-10 keV luminosities from the 2-10 keV
fluxes, when available, and from the 0.5-2 keV fluxes in the remaining
cases. We assumed a power law spectrum with
$\alpha_E=0.8$ reduced at low energy by rest frame photoelectric
absorption.  We used the N$_H$ values reported by Tozzi et al. (2006)
for the sources with redshift consistent between this and the previous
work ($\sim 68$\% of the sources have $|\Delta z/(1+z)|<0.2$, see also
next section). For the remaining 32\% of the sources we evaluated N$_H$ 
from the hardness ratio between the hard and soft bands.

%                                     Two column figure (place early!)
%______________________________________________ Gamma_1 (lg rho, lg e)
   \begin{figure}[!t]
 \includegraphics[width=8.7cm]{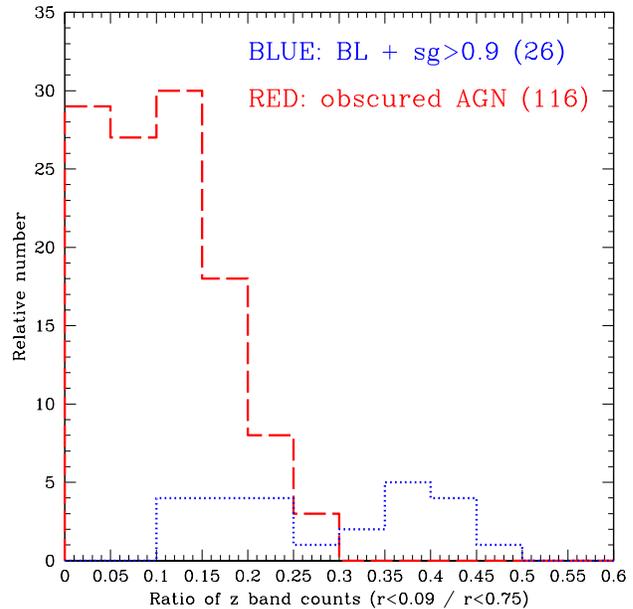}
   \caption{Distribution of the $z-$band count ratio for the
``obscured'' AGN sample (116 objects, red long dashed histogram) 
and the BL/pointlike sample (26 objects, blue dotted histogram). }
   \label{ratio}%
    \end{figure}
\subsection{Optical+Infrared photometry and photometric redshifts}

Table 2 shows the compilation of optical b,v,i,z bands (col 2-5) from
GOODS data (Santini et al. 2009), infrared J,H,K (col. 6-8) from
ESO/VLT, IRAC and MIPS (col. 9-13) photometry.  Of the 179 X--ray
sources, 174 are detected in the z-band, 172 in the 4.5$\mu$m IRAC
channel and 133 in the 24$\mu$m MIPS survey. \\ The spectroscopic or
photometric redshifts available for the proposed optical counterparts
of the 179 CDFS-GOODS sources are listed in column 10 of Table 1.  When a
spectroscopic redshift is present, the relative source catalog is
reported, as follows: (1) CXO-CDFS (S04); (2) K20 (Mignoli et
al. 2005); (3) GOODS (Vanzella et al. 2005, 2006, 2008, Popesso et
al. 2008); (4) COMBO-17 (Wolf et al. 2003); (5) VVDS (Le Fevre et al. 2004); 
(6) Other (Daddi et al. 2005, Cristiani et al. 2000).  Otherwise, the
redshift is from SED fitting (Santini et al. 2009) and is reported
with its 1 $\sigma$ confidence interval.  We have spectroscopic or
photometric redshifts for all but one (AID 282) of the identified
X-ray sources.  However, in eleven cases, only a lower limit on the
photometric redshift solution is available from the SED fitting,
i.e. at higher redshifts the $\chi^2$ curve is relatively flat
and an upper limit on the basis of a $\Delta\chi^2$ could not be
determined. 
 \\
Figure 2 shows the redshift distribution of the entire sample (open
black histogram) and the distributions derived for the spectroscopic
and photometric sub-samples (blue and red histograms,
respectively). About half of the redshifts at z$>1$ come from the
photometric sample. For about 25\% of the sources the redshifts listed
in Table 2 are different (i.e. not consistent within the errors)
from the ones published in previous works (as compiled by Zheng
et al. 2004, Tozzi et al. 2006), for the following reasons: 1) for
9 objects the counterparts have changed from the ones reported in 
the literature (see previous section); 2) for about 30 sources 
the availability of  longer wavelength (IRAC) data increases the 
reliability of the photometric redshifts.
The photometric redshifts for the sources in the MUSIC
sample are instead consistent with a new analysis of the sources in
the 2 Msec CDFS (Luo et al. 2008) and the Extended Chandra Deep Field
South (E-CDFS, Lehmer et al. 2005) surveys obtained by Rafferty et
al. (2009, in preparation). \\

%----------------------------------------------------------- S_vib
   \begin{figure}[!t]
   \includegraphics[width=9cm]{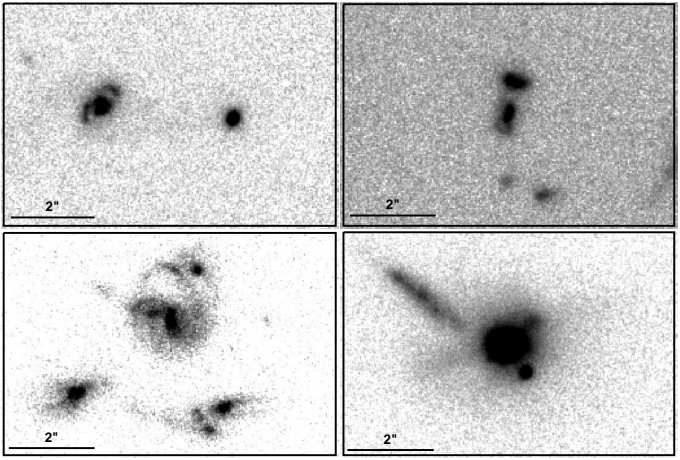}
      \caption{ACS ($z-$band) images of obscured AGN counterparts with 
disturbed morphology or interacting systems (AID 88, 107, 226 and 84).}
         \label{FigVibStab}
   \end{figure}

%______________________________________________ Gamma_1 (lg rho, lg e)
   \begin{figure*}[!t]
\includegraphics[width=8.7cm]{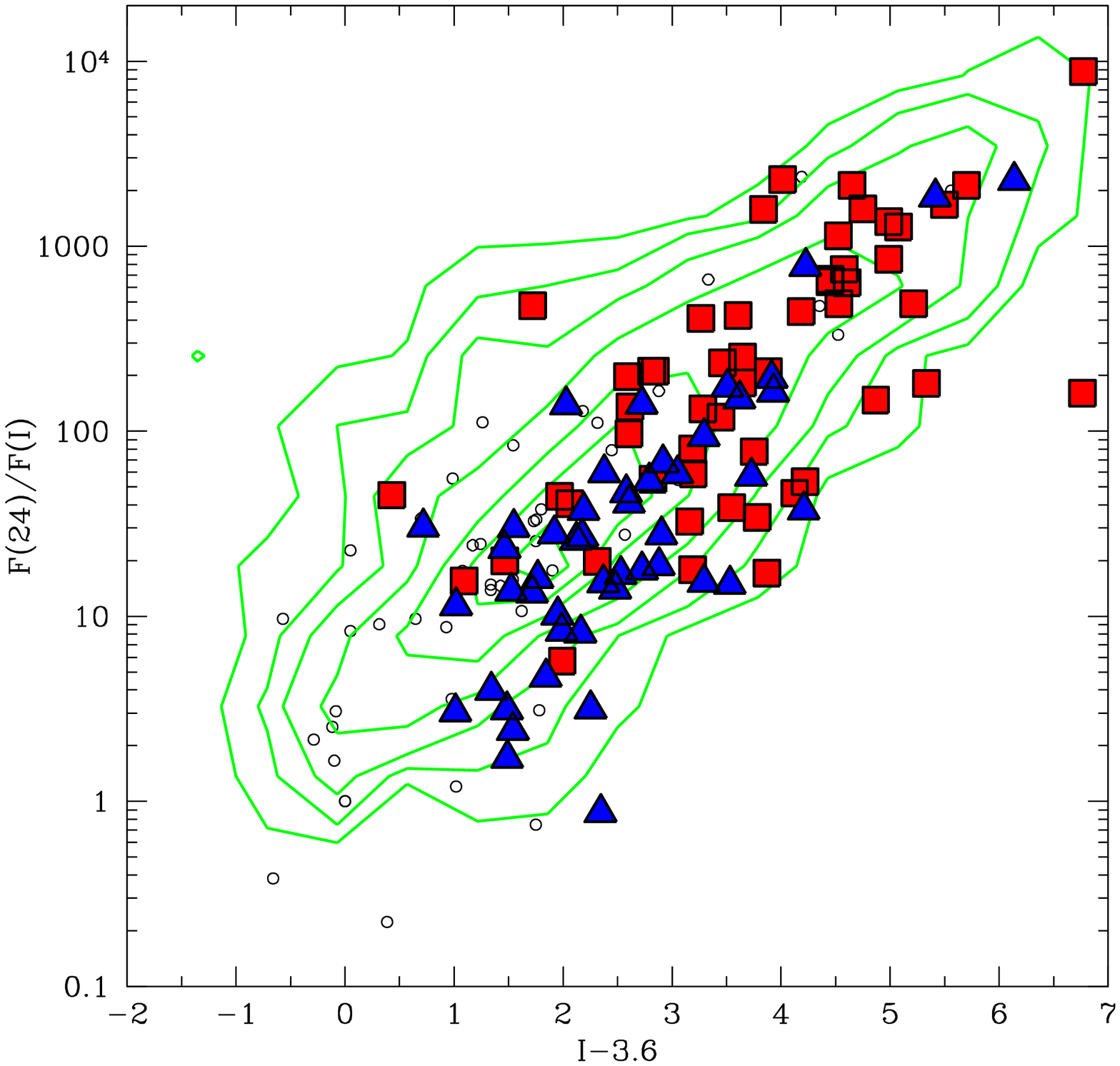} 
 \includegraphics[width=8.7cm]{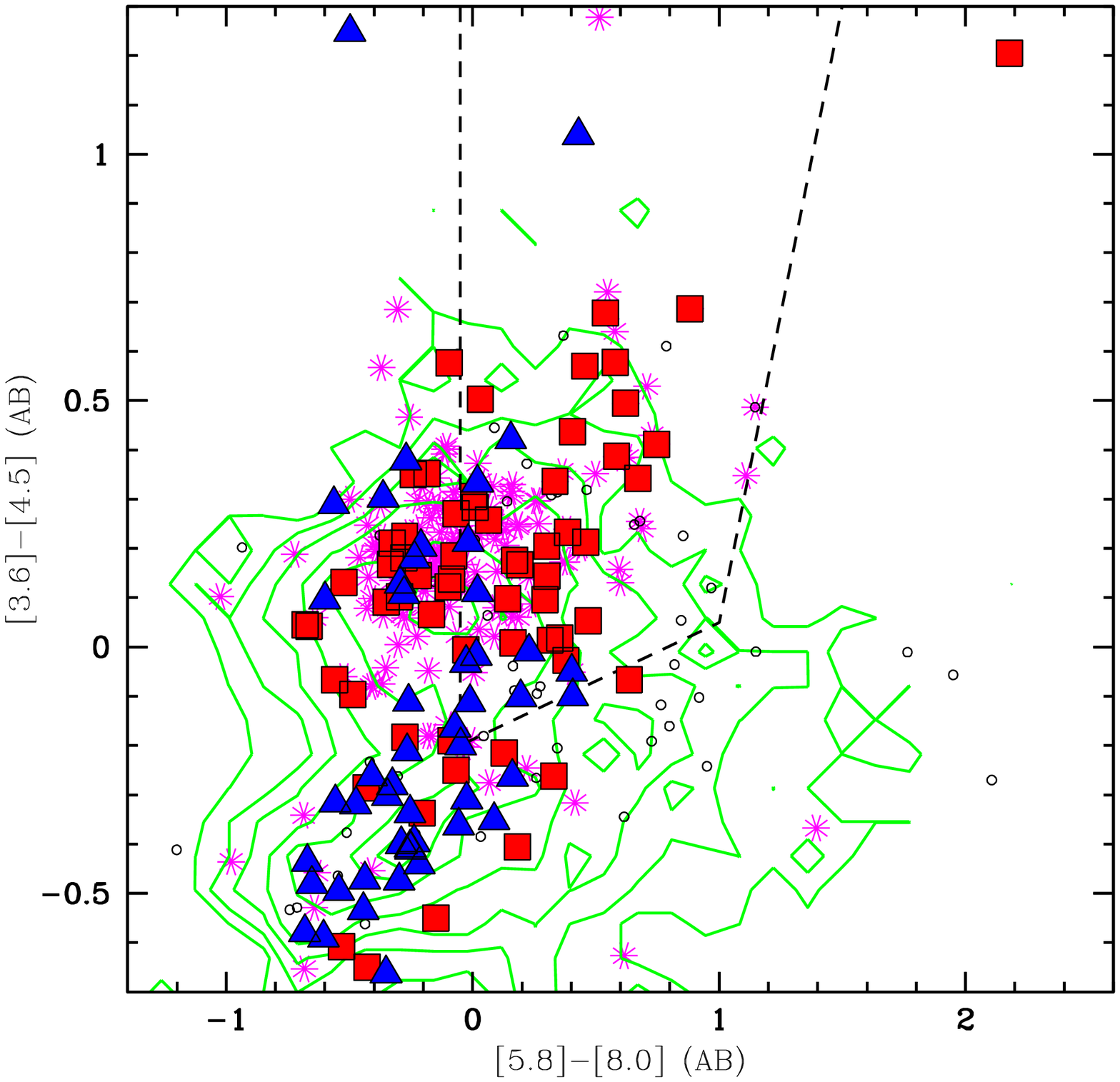}
   \caption{
{\it Left Panel}: 
24 $\mu$m to optical (I-band) flux ratio vs. the I-[3.6] color for the obscured
AGN in our sample.
The sizes and shape (colors) of the symbols reflect the AGN X--ray luminosity:
(blue) small triangles are objects with logL$_{X}<43$, (red) large squares 
objects with  logL$_{X}>43$. Open circles mark all the other X-ray sources 
(both BL/pointlike AGN and sources with logL(X)$<$41.8). 
The green contours are the isodensities of the 24 $\mu$m selected population.
{\it Right Panel}: 
IRAC color-color: [3.6]-[4.5] color vs. [5.8]-[8.0] color.  
The dotted lines isolate a wedge for the selection of bright, optically
unobscured quasars (see Stern et al. 2005). 
The color-code is the same as in left panel. The magenta asterisks 
are objects in the MIPS sample selected on the basis of their high 
24$\mu$m to optical flux ratio and extremely red colors (R-K$>$4.5).}
   \label{FigGam}
    \end{figure*}

%                                     Two column figure (place early!)

\subsection{Obscured AGN sample}
 
We extracted from our sample of 179 sources, the sample of
``obscured'' AGN defined as outlined below.  First, we considered all
the sources with rest frame logL(2-10 keV)$>41.8$ in order to
minimize the contribution of non-AGN objects among the X-ray selected
sample. Indeed, following Ranalli et al. (2003), the observed X--ray
emission can be due to stellar processes only in objects with SFR larger than 
$\sim$100 M$\odot$ yr$^{-1}$ (see also Section 4.2 and 4.4 for further 
discussion on the luminosity threshold).  
We also removed one source that shows extended X-ray emission,
and the three sources without a unique/secure identification (see Section 2.1).
We were left with 142 sources.  
Secondly, we isolated all the 19 sources
spectroscopically identified as Broad Line (BL) AGN in Szokoly et al. (2004),
and additional 7 objects still spectroscopically unidentified but with 
a {\it sg} parameter (stellarity, from SExtractor, Bertin \& Arnouts 1996) 
larger than 0.9. We named these 26 sources as BL/pointlike sample.  
Following Silverman et al. (2008), we
computed the ratio of the flux in the $z_{850}-$band in two fixed
apertures (a small aperture with a radius of 3 pixels (0.09'') and a
larger aperture with a radius of 25 pixels (0.75'')). This quantity is
similar to the inverse of the concentration index (Abraham et
al. 1994). 
Fluxes were determined by extracting counts from background-subtracted
cutouts of ACS images centered aroud the position of the X--ray sources.
Figure 3 shows the distribution of the $z-$band counts ratio for
the BL/pointlike sample (blue histogram), and the remaining sources (116
objects, red histogram).  All the objects with ratio values $>0.3$
are from the BL/pointlike sample, as expected in sources for which the
pointlike and/or nuclear emission dominates overall the host-galaxy
light.  The contribution of pointlike/unobscured AGN at ratio values
$<0.3$ is $\sim10$\% (being zero in the two lowest ratio bins, 0-0.1),
therefore we expect a negligible contamination from this class of
sources in the sample with ratio $<0.3$.  We call ``obscured'' AGN the
116 objects with the ratio of z-band counts $<0.3$ and not classified
as BL AGN or pointlike in the ACS images. 
Ninety of these "obscured" AGN ($\sim 75$\%) are also detected
at MIPS 24$\mu$m. \\

X--ray column densities from the 1 Ms sources in the CDFS derived from
an accurate spectral fitting analysis have been published by Tozzi et
al. (2006). As discussed in the previous Section, a fraction of 
the X--ray to optical associations changed from the Tozzi et al. 
(2006) analysis to the present analysis (9 sources), and in addition some more
robust photometric redshifts have been obtained using the Spitzer
photometry.  Among the sixty-nine sources in the ``obscured'' AGN
sample in common with the Tozzi et al. sample\footnote{We recall that
our sample contains 16 sources detected only in the A03
 sample and not present at all in the Tozzi et al. (2006)
spectral analysis} and with a $|\Delta z/(1+z)|<0.2$ from the redshift
values reported in Tozzi et al. (2006), fifty-four ($\sim 80$\%) show
X--ray absorption in excess of 10$^{22}$ cm$^{-2}$, further suggesting
that our sample is representative of the X--ray obscured AGN population.

We visually inspected all the sources in the z-band ACS images and
divided them in three main classes: isolated sources (35),
interacting/disturbed/clumpy galaxies (53) and objects too faint to
tell something on the morphology (26).  All of these 26 sources
populate the first bins in the histogram in Fig.~3.\\ Figure~4 shows
four examples of obscured AGN counterparts with disturbed morphology
or interacting systems.

\subsection{Host galaxy properties}
For the host galaxies of the obscured AGN we retrieved from the
GOODS-MUSIC catalog the rest frame U-V colors, the absolute V and
K-band magnitudes, the SFRs and the stellar masses M$_*$ derived from the
SED fitting. 
More specifically, 
stellar masses are derived from the comparison of SED 
and libraries of synthetic spectra, assuming a Salpeter (1955) Initial
Mass Function (IMF) and 
simple stellar population (SPP) models (Bruzual \& Charlot 2003), 
as described at length in Fontana et al. (2006).
SFRs have been estimated from SED fitting, 
using Bruzual \& Charlot (2003) synthetic models and fitting the 
UV rest frame photometric bands. 
We parametrize
the star formation histories with a variety of exponentially declining
laws (with timescales  ranging from 0.1 to 15 Gyr),
metallicities (from Z = 0.02 Z$\odot$ to Z = 2.5 Z$\odot$) 
and dust extinctions (0$ < $E(B-V) $< 1.1$, with a Calzetti or Small Magellanic
extinction curve). These SFR estimates have been compared with the ones
derived from the observed IR (MIPS) emission, following the approach 
presented by Papovich et al. (2007): despite the majority of galaxies 
are assigned a consistent SFR (with overall median ratio around unity,  
see details in Santini et al. 2009), in the following, we will use 
the SFR from the SED fitting for all the obscured AGN in our sample, 
in order to minimize the possible AGN contribution in the computation 
of SFRs from MIPS data in AGN samples (see Daddi et al. 2007a).  \\
As a word of caution, we note here that the derivation of host 
galaxy properties is affected by both statistical and systematics 
undertainties.  \\ 
Statistical errors for both masses and SFR have been calculated 
from the 1$\sigma$ convidence level in each parameter estimate by scanning
the $\chi^2$ levels, allowing the redshift to change within the errors 
in case of objects with photometric redshifts\footnote{We note here
that photo-z accuracy, in general, is not the leading source of error in 
deriving the galaxies and AGN parameters (see also discussion in Ilbert 
et al. 2009).} (see Fontana et al. 2006, Santini et al. 2009). \\ 
The masses and SFR estimates are, however, limited by the systematics 
related to model degeneracy and, in particular, the ones associated 
to the adopted IMF, the SSP, the metallicity and the star formation 
history (see Santini et al. 2009 and the discussion in Ilbert et al. 2009). 
The largest, systematic, uncertainty on the estimate of the stellar mass 
is the assumption on the IMF. 
The masses obtained using a Chabrier IMF are on average 0.24 dex lower 
than those obtained adopting a Salpeter IMF, and this effect is totally 
independent from the range of redshift and mass probed (Salimbeni et al.
2009). 
The uncertainties related to the adopted SSP models (Bruzual \& 
Charlot 2003, Charlot \& Bruzual 2007, Maraston 2005) to construct 
libraries of synthetic spectra, are also of the order of 
Delta (logM) = 0.2-0.3, but the effect is less evident in the high-mass 
part and at high-z (see Salimbeni et al. 2009 for a proper discussion 
on these effects in the GOODS-MUSIC sample). \\ 
Similar systematics effects are present also in the SFR estimates,
and are expected to be of the order of Delta(logSFR)$\sim0.2-0.4$
(see, e.g., discussion in Ferreras et al. 2005 for the 
variation due to different IMF).  \\
Since most of these effects are systematics
and since we are interested at general trends over several decades 
of mass and not in the detailed description of the mass of the host 
galaxies of each X-ray source, we are confident that the overall 
results discussed in the rest of the paper remain unchanged.

\section{Observed frame colors} 
A proper classification of source nature (AGN vs. starburst vs. normal
evolved galaxy) should ideally be obtained via a complete analysis
of its emission over the electromagnetic spectrum, using both
spectroscopic (e.g. emission line widths and ratios) and photometric
(all-wavelengths sampled SED) observables.
However, in reality, a complete source characterization is
not obtainable, even in fields where the best and deepest imaging
and spectroscopic campaigns have been performed, such as the GOODS/MUSIC
field. Partial, but still reliable, information (at least in a statistical sense)
can be obtained through the analysis of
the emission in bands where differences between
nuclear and star formation emission are emphasized, comparing
the observed SEDs with galaxy templates over a range as broad
as possible.
In particular, the combination of observed-frame mid-infrared, near infrared and optical flux ratios
has been exploited recently in the literature to isolate obscured AGN
(see, e.g., Martinez-Sansigre et al. 2005, 2006,
Fiore et al. 2008, Donley et al. 2008, Dey et al. 2008).  \\
\par\noindent
Figure 5, left panel
shows the 24$\mu$m to optical (I-band) flux ratio vs. the I-[3.6]
color for the X-ray sources in our sample.  The green contours show
the isodensities of the field objects in the MIPS selected sample.
The sizes and shape (colors) of the symbols reflect the AGN 2-10 keV
luminosity: 
(blue) small triangles are objects with logL$_{X}<43$, (red) large squares 
objects with  logL$_{X}>43$. Open circles mark all the other X-ray sources 
(both BL/pointlike AGN and sources with logL(X)$<$41.8). 
The well-established correlation between the MIR/O
color and the total source luminosity traced in this case by the 
X--ray emission (see Fiore et al. 2008) is also present in this plot, 
where the most luminous sources (red squares)
cluster at average values of MIR/O ratio higher than those of less
luminous sources (blue triangles), and on average towards redder optical
to near infrared colors. \\
The right panel of Figure 5 shows an IRAC color-color diagram where the
[3.6]-[4.5] color is plotted against the [5.8]-[8.0] color.  The
dotted lines isolate a wedge for the selection of bright, optically
unobscured quasars, originally proposed in Stern et al. (2005).  The
sizes and shapes (colors) of the symbols reflect the AGN X-ray luminosity, as
in the left panel of Fig. 5.  Open small symbols are unobscured AGN
and all the other X--ray sources.
While the most luminous "obscured" AGN occupy the region in the
color-color diagram expected from a power-law SED, i.e.within the
wedge marked by the dashed lines, the moderate luminous "obscured" AGN
population mainly clusters in two regions with roughly the same
[5.8]-[8.0] color but with a bimodal distribution in the [3.6]-[4.5] color
(see also results presented by Cardamone et al. 2008) .
About half of the moderate luminous obscured AGN in the sample are
clustered around [3.6]-[4.5]$\sim-0.25$. Most of these objects are in
the redshift range z=0.6-0.8 (i.e. in the structures present in the CDFS field,
see Gilli et al. 2005, Silverman et al. 2008).  
In this region also fall the 60\% (4/7) of the radio detected sources
in the obscured AGN sample, extracted from the work by Tozzi et al. (2009,
see also the results presented by Hickox et al. 2009) 
The other half of the obscured AGN sample clusters at redder 
[3.6]-[4.5] colors ($\sim0.25$), where both
star-forming and normal galaxies at z$\sim 1-3$ are expected.  The
isodensity curves from the GOODS-MUSIC sample (Santini et al. 2009),
overlaid as green contours in the plot, show the same bimodal
distribution as observed for the X-ray sources. \\ 
In Figure 5, right
panel, we also plot as magenta asterisks the population of the most
obscured, possibly Compton Thick AGN at z$\sim2$ discussed by Fiore et
al. (2008) and selected on the basis of their high 24$\mu$m to optical
flux ratio and extremely red colors (R-K$>$4.5; see also Daddi et
al. 2007a, Dey et al. 2008). 
These X--ray undetected, Compton Thick AGN candidates
occupy the same region in which X--ray detected, $z>1$ obscured AGN
lie (see also Georgantopoulos et al. 2008). \\

%
%                                     Two column figure (place early!)
%______________________________________________ Gamma_1 (lg rho, lg e)
   \begin{figure}[!t] 
\includegraphics[width=8.7cm]{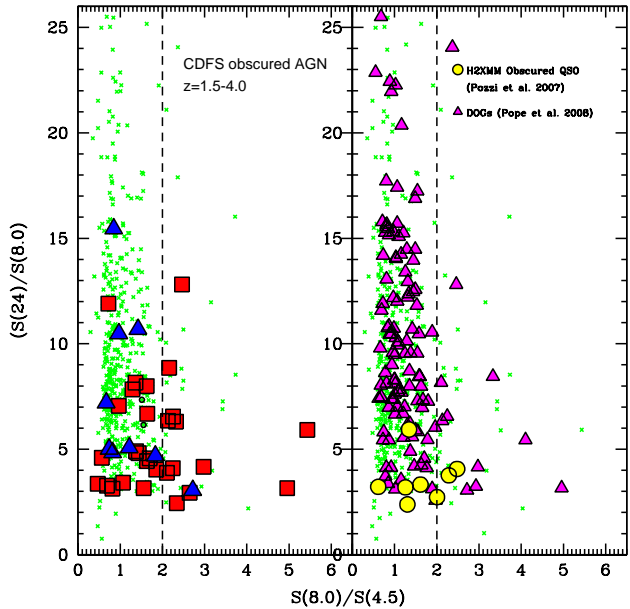}
   \caption{Spitzer color color diagram used to isolate star-formation and AGN
   dominated high-z galaxies (Pope et al. 2008). The small green points 
   represent the distribution of
   field galaxies in the 24$\mu$m MUSIC sample in the same redshift range. Objects with
   S(8.0)/S(4.5) $<2$ are classified ``star-formation dominated'' 
   by Pope et al. (2008). In the left panel we superimpose to the MUSIC/MIPS
   population the objects in our obscured AGN sample. The color code for
   the large symbols is the same as in Fig. 5. Only objects at z=1.5-4
   have been included. In the right panel other samples of high-z obscured
   AGN are plotted, as labeled.}  
\label{pope}% 
\end{figure}

Figure 6 shows the combined MIPS-IRAC color-color diagram introduced by
Pope et al. (2008, see also Ivison et al. 2004) to distinguish between
AGN and star-formation dominated sources in high-z (z$>1.5$) infrared selected
sources, where the 24$\mu$m to 8.0$\mu$m flux ratio is plotted against the 8.0$\mu$m to 4.5$\mu$m
flux ratio. Indeed, at z$>1.5$, the observed 24$\mu$m emission 
samples the rest frame $<10\mu$m emission, where strong starburst PAH
lines are commonly present and can boost the measured IR flux,
giving a higher 24$\mu$m to 8.0$\mu$m flux ratio. Similarly, the 
8.0$\mu$m to 4.5$\mu$m flux ratio is a measure of the rest-frame 
near infrared continuum slope.
Pope et al. (2008) proposed that dust obscured galaxies 
with  large values of the 8.0$\mu$m to 4.5$\mu$m flux ratio ($>2$; dotted line 
in Fig. 6) are AGN dominated sources, according to the presence of AGN 
continuum in Spitzer IRS spectra, while objects with lower values of the 
8.0$\mu$m to 4.5$\mu$m flux ratio ($<2$) are ``star-formation dominated'',
in agreement with the fact that they span a wide range of 
24$\mu$m to 8.0$\mu$m flux ratio.
In both panels of this figure, we plot the field 
population extracted from the MUSIC sample as small crosses. 
In the left panel, we report as squares the CDFS  obscured AGN with a 
MIPS 24$\mu$m  detection, and in the redshift range z=1.5-4 (38 objects 
in total).  
In the right panel, we mark as triangles the sources satisfying, among
the underlying population, the dust obscured galaxies selection used 
by Pope et al. (MIPS/O ratio larger than 1000 and R-K$>4.5$, see also 
Fiore et al. 2008, Dey et al. 2008)
and as circles the X--ray selected obscured QSO from the Hellas2XMM survey
for which deep Spitzer imaging has been obtained (Pozzi et al. 2007).\\
According to the classification introduced by Pope et al. (2008), 
the subsample of X--ray selected, "obscured" AGN with MIPS detection show 
Spitzer colors consistent with both an AGN dominated continuum and a starburst 
dominated continuum in the mid-infrared. Actually, a large fraction of the most X-ray luminous AGN
lie in the Pope et al. (2008) ``star-formation dominated'' part of the
diagram, similarly to the X-ray selected obscured quasars from
the Hellas2XMM survey. 
In conclusion, for at least half of the obscured AGN with MIPS detection 
in our sample, the accretion activity is revealed thanks to the presence of the
X--ray emission and not by a large 8.0$\mu$m to 4.5$\mu$m flux ratio.
Moreover, about half (13/25) of the sources classified as AGN in the 
"starburst" part of the diagram have 24$\mu$m to 8.0$\mu$m flux ratio 
lower than 5, while in the original Pope et al. diagram this region
is almost empty.

\section{Rest-frame colors: Host galaxy properties}

\subsection{Reddening}

%                                     Two column figure (place early!)
%______________________________________________ Gamma_1 (lg rho, lg e)
   \begin{figure*}[!th]
 \includegraphics[width=8.7cm]{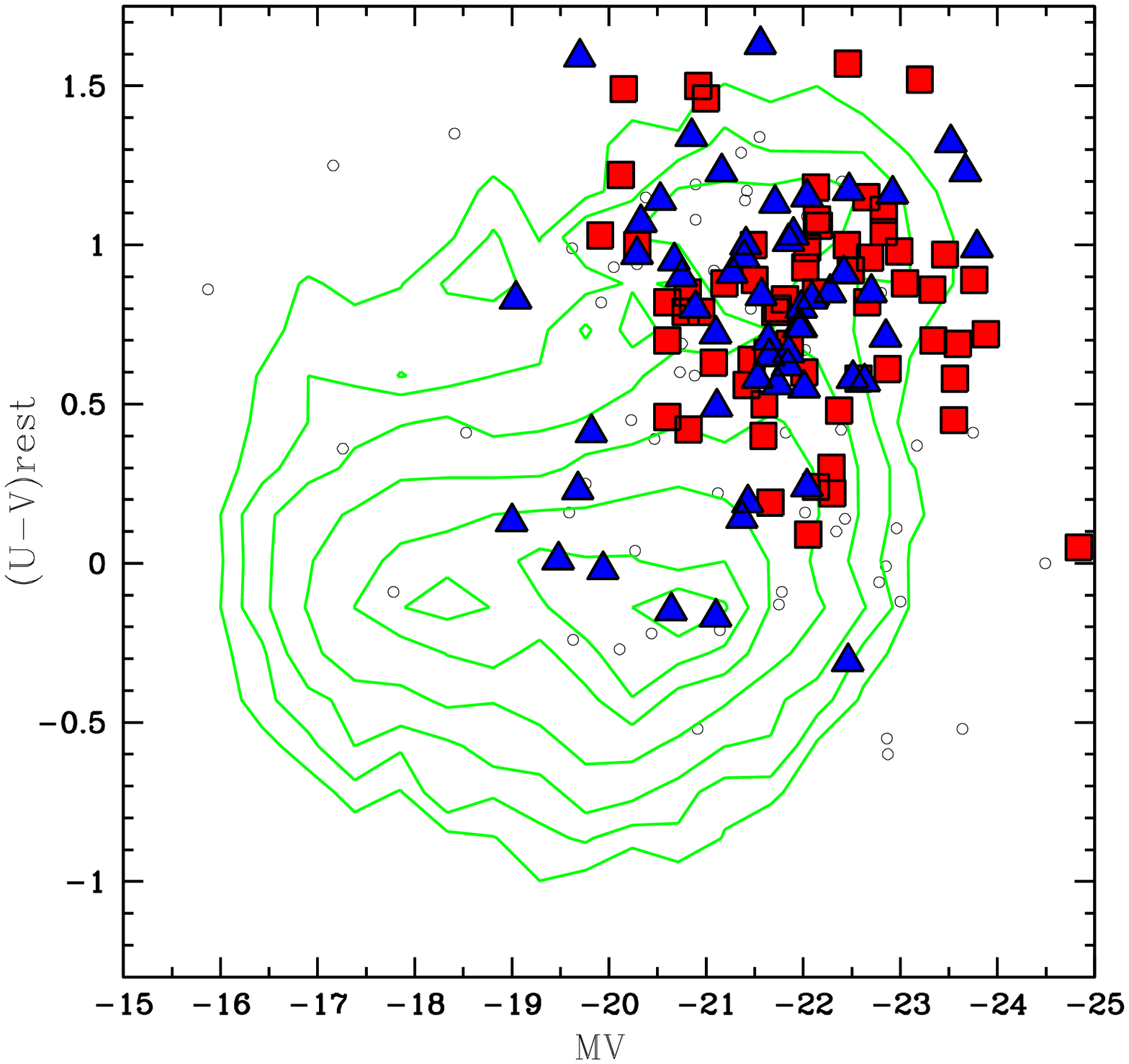}	
 \includegraphics[width=9cm]{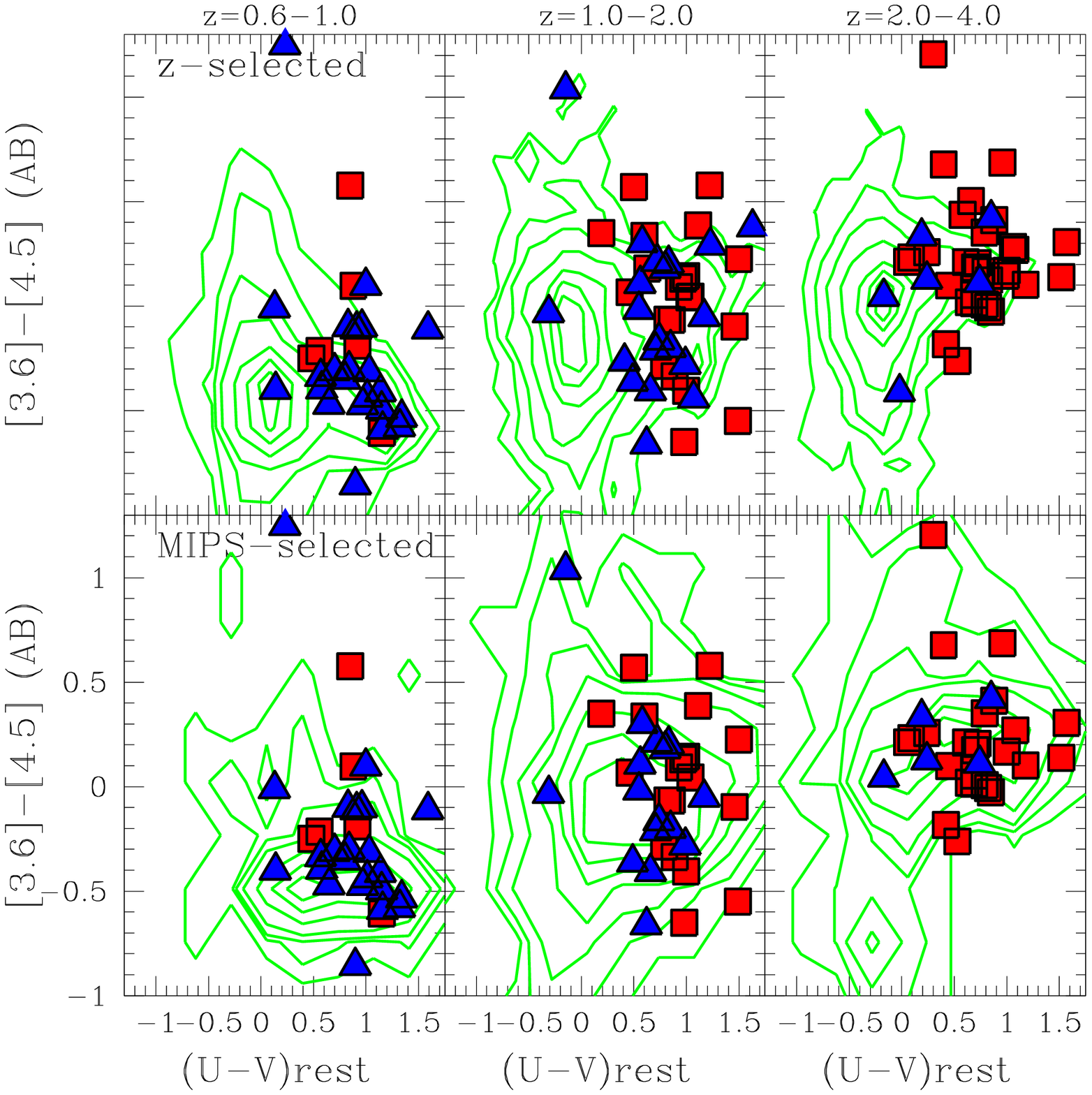}

   \caption{
{\it Left Panel:} U-V(rest frame) vs. the absolute V-band magnitude for X-ray obscured AGN
     (squares) and the underlying optically selected galaxy population (contours).
     The sizes and colors of the symbols reflect the AGN X--ray luminosity
     (as in Fig. 5 and 6). 
 {\it Right Panel:} [3.6]-[4.5] color vs. the U-V(rest frame) for obscured X-ray sources
     (squares) and the underlying optically selected galaxy population (contours). The upper
     panels show the comparison of the X--ray selected sources with the
     optically-selected (z-band) galaxy population in three different redshift
     bins: z=0.6-1.0, z=1-2, and z=2.4 from left to right, respectively. The
     lower panels show the comparison for the subsample of objects detected
     also at 24 micron, in the same redshift bins. The sizes and colors of the 
symbols reflect the AGN X-ray luminosity as in Fig. 5 and 6. Unobscured AGN are 
not shown here.}
   \label{FigGam}
    \end{figure*}

The left panel of Figure 7 shows the U-V(rest) vs. the absolute V-band
magnitude for X-ray selected "obscured" AGN (squares) and the underlying 
optically selected galaxy
population (contours) in the redshift interval z=0.6-4.0. 
The sizes and colors of the symbols reflect the AGN X--ray luminosity
(as in Fig. 5 and 6). 
X--ray selected, obscured AGN show a (U-V) rest color
redder than the average color of the optically selected galaxy
population and are hosted preferentially by luminous (MV$<-20$), red
(U-V$>0.5$) galaxies, populating mostly the red sequence and the green
valley of the bimodal color-magnitude distribution observed for field
objects. We do not find any trend with the X-ray luminosity: the U-V color does 
not depend on the strength of the X--ray emission. 
This result is in agreement with previous findings reported
in the literature (Nandra et al. 2007, Rovilos \& Georgantoupolos 2007, 
Silverman et al. 2008, Hasinger 2008); in the next Section
we will further discuss it and its consequences for our
fully understanding of AGN hosts properties.

The right panel of Figure 7 shows the [3.6]-[4.5] color vs. the U-V (rest
frame) color for the sample of obscured AGN (squares) and the
underlying optically selected galaxy population (contours), in three different redshift
bins: z=0.6-1.0, z=1-2, and z=2-4 from left to right, respectively.
The upper panels show the comparison of X--ray selected obscured
sources with the optically-selected (z-band) galaxy population; the
lower panels show the comparison for the subsample of objects detected
also at 24$\mu$m.   \\ 
We do not find a significant evolution in the
(U-V) rest frame color with redshift for both the overall galaxy population and
the obscured AGN sample. Instead, the average value of the [3.6]-[4.5]
color increases with redshift and 
the average [3.6]-[4.5] color of the obscured AGN is consistent 
with that of the underlying galaxy population (see also Fig. 5).  The
only difference among the field and obscured AGN population is,
therefore, the fact that obscured AGN are redder, in U-V color,  
than optically selected sources (as outlined in the left panel), while sharing
the same [3.6]-[4.5] color. This is probably due to the fact that the
observed [3.6]-[4.5] color strongly correlates with redshift and it is not
related to any intrinsic property of the sources (e.g. star formation
or AGN activity).  
If we consider the MIPS selected sample (lower panels) the
X--ray selected, obscured AGN closely follow the distributions of the
underlying star-forming population with redder U-V rest frame
colors than in the z-selected sample.  \\

%                                     Two column figure (place early!)
%______________________________________________ Gamma_1 (lg rho, lg e)
   \begin{figure*}[!t]
 \includegraphics[width=9cm]{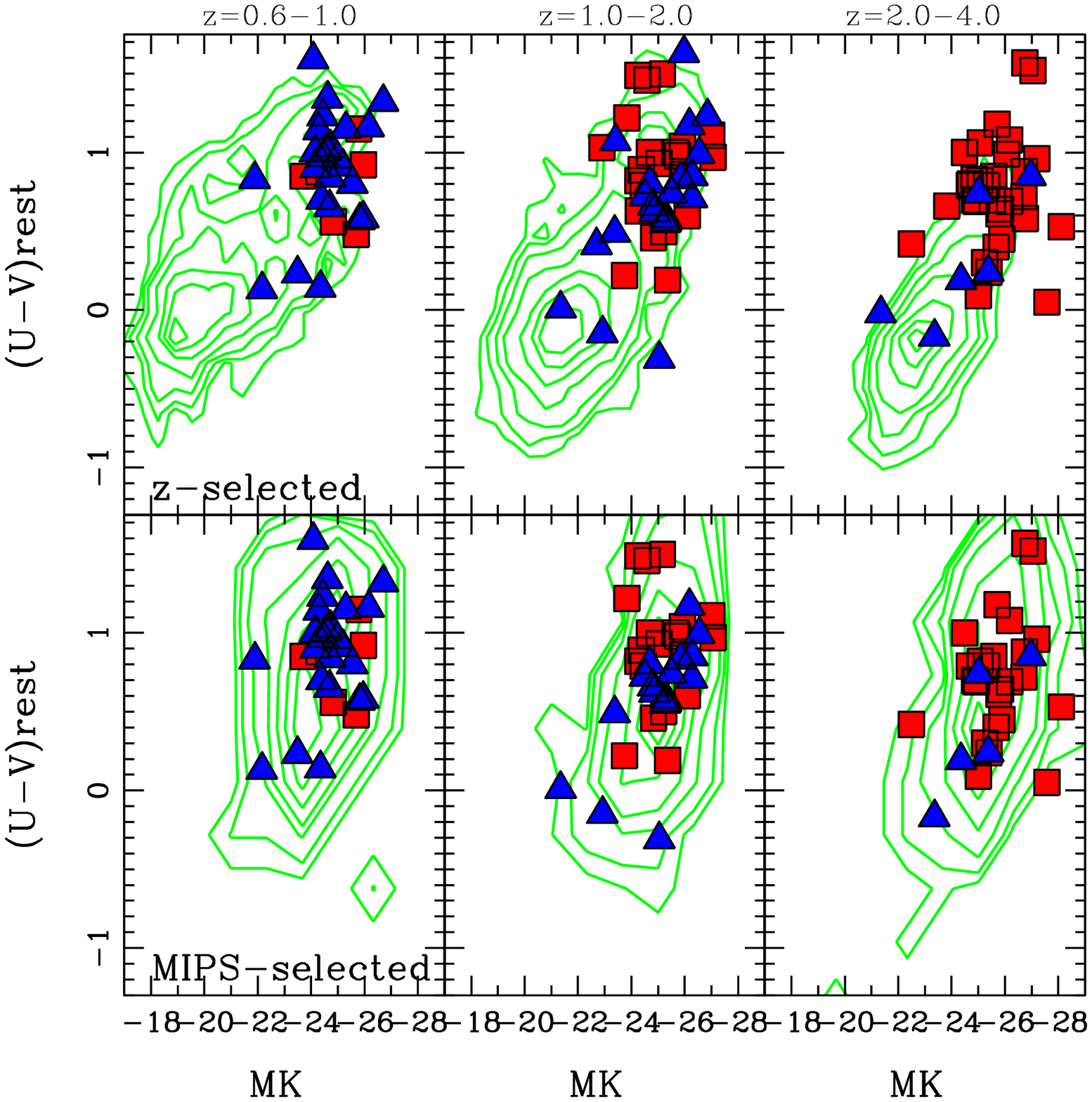}
 \includegraphics[width=9cm]{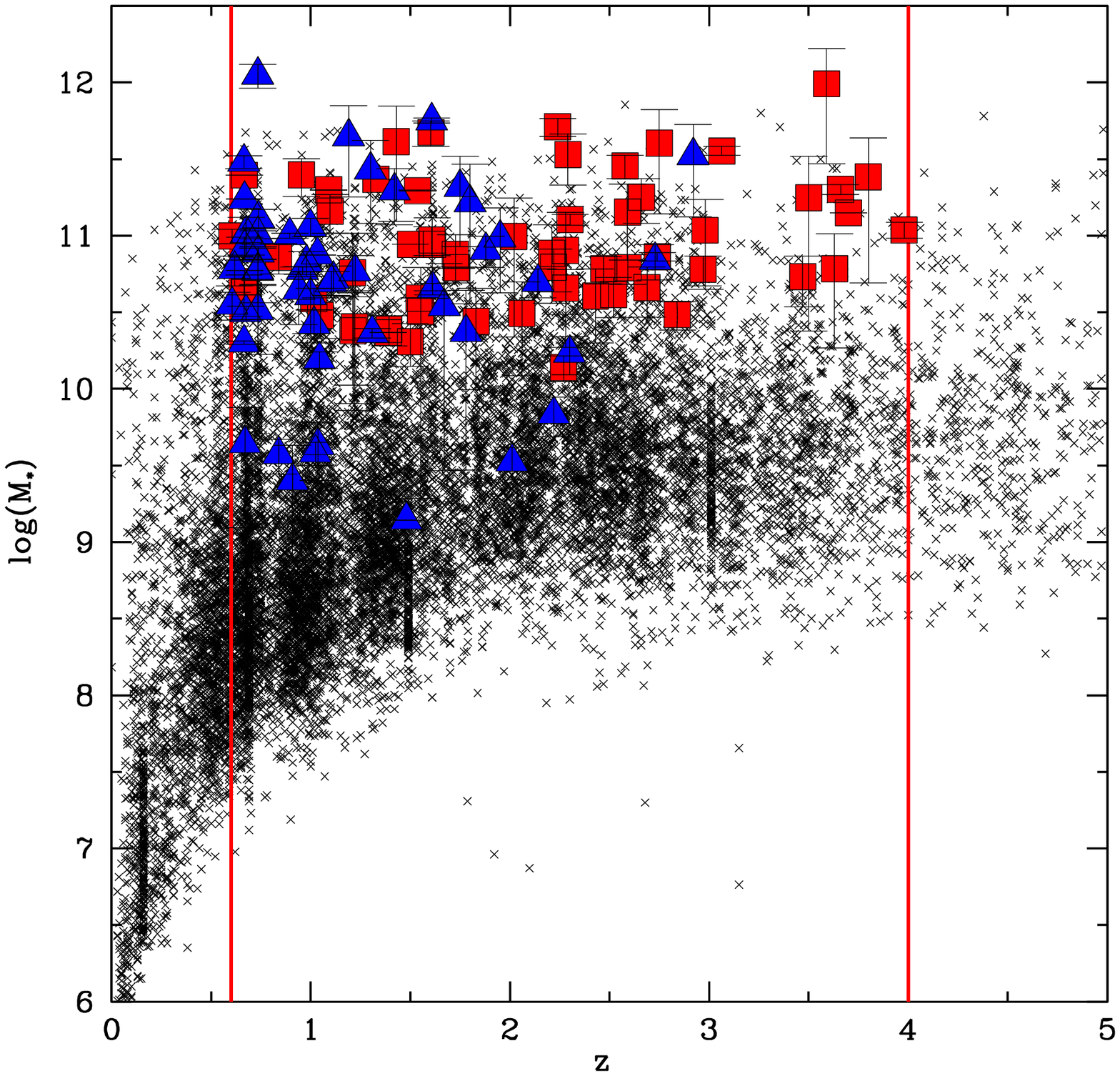}
   \caption{
{\it Left Panel:} U-V(rest frame) vs. the absolute K-band magnitude for the obscured AGN 
     (squares) and the underlying galaxy population (contours). The upper
     panels show the comparison of the X--ray selected sources with the
     optically-selected (z-band) galaxy population in three different redshift
     bins: z=0.6-1.0, z=1-2, and z=2.4 from left to right, respectively. The
     lower panels show the comparison for the subsample of objects detected
     also at 24 micron, in the same redshift bins.
     The shapes and colors of the symbols reflect the AGN X--ray luminosity
(see Fig. 5). 
{\it Right 
Panel:} Stellar mass M$_*$ vs. redshift for field galaxies
(crosses) and our ``obscured'' AGN (squares and circles, same notation as in Figure 5).  
The vertical lines mark the redshift range we chose for our analysis. }
   \label{FigGam}%
    \end{figure*}

Figure 8 (left panel), shows the U-V(rest frame) vs. the absolute K-band magnitude
for the obscured AGN in our sample (squares) and the isodensity contours of the galaxy
population (green curves), in the same redshift bins as in Fig. 7,
and for both the optically (z-band) and MIPS selected samples. 
\\ At all redshifts, most of the obscured AGN host galaxies show absolute
magnitude brighter than M$_{K}<-23$, much brighter than that of the bulk of
the optically selected population (upper panel): the large majority of
optically selected galaxies cluster around lower values of M$_{K}$ and
bluer values of the (U-V) rest frame colors. \\
The distribution of sources in the U-V vs M$_{K}$ plane is different when
the MIPS selected sample is considered: obscured AGN hosts share
the same U-V colors of the mid infrared selected galaxy population. 
In this case, a color
bimodality is not observed at any redshift, and the U-V color
correlates smoothly with the galaxy mass (also see Aussel et
al. 2009). AGN hosts are found preferably at the peak of the galaxy
isodensity contours in Fig. 8 lower panels.  
In particular, 
the comparison of the upper and lower panels of Fig. 8 suggests a
scenario in which star forming galaxies hosted in low mass systems are
not particularly dust obscured and therefore the reprocessed emission
is not revealed at IR bands (at least at the limiting fluxes of the
24$\mu$m GOODS survey), consistent with the results by Daddi et al. 2007b. \\
\par\noindent
Given that M$_{K}$ is a good tracer of the
stellar mass of the galaxies, this suggests that obscured
AGN are preferentially hosted in high-mass systems.  This is shown in
the right panel of Fig. 8, where the best fit logM$_*$ is plotted against the
redshift (see also Alonso-Herrero et al. 2008) for both the field
population from the MUSIC catalog (crosses) and the obscured AGN (red
circles). As discussed in Section 2.4, the determination of the stellar 
mass suffers from severe systematic uncertainties. However, these 
uncertainties are present for both the AGN and the underlying galaxy
population, with the net results of a likely shift in the Y-axis 
($\Delta$logM$_*$=-0.3). \\

\subsection{Star formation activity}

The black open histogram in Figure 9 left panel, shows a normalized 
distribution of the SFR (see Section 2.4) for both the entire MUSIC 
catalog (upper panel) and the MIPS selected sample (lower panel).  
The red filled histogram shows the distribution
of SFR for the ``obscured'' AGN in our sample. In all cases the SFR
have been restricted to objects in the redshift interval z=0.6-4, and
in the stellar mass range M$_*=10^{10}-10^{12}$ M$_\odot$.  
We chose this stellar mass range because most of the obscured 
AGN in our sample have masses larger than this value (see Figure 8
right panel) and the completeness of both the entire MUSIC catalog and 
the MIPS selected sub-sample have been studied in details 
(see discussion in Fontana and Santini et al. 2009). 
The total sample considered here includes 103 obscured
AGN (82 in the MIPS detected subsample). 
\\ 
In $\sim 65$\% of the cases the host galaxies of these AGN
show substantial ($>10$ M$_\odot$ yr$^{-1}$) amount of SFR. 
The distribution of SFR is significantly different from the one of optically
selected galaxies (the probability that the two distributions shown in the 
upper-left panel of Figure 9 are drawn from 
the same parent population is only 0.1\%, according to a Kolmogorov-Smirnov 
statistics) suggesting an enhancement of
star formation activity in objects hosting obscured AGN. 
This result is solid despite the large systematics uncertainties 
(dominating the SFR measurements, see discussion in Section 2.4), 
given that they are expected to  be present in {\it both} the optically 
and X-ray selected population. \\
An enhancement of the SFR activity in obscured AGN  is qualitatively 
in agreement with the results at z$\sim0.7$ obtained
from an X--ray selected sample of AGN from the XMM-COSMOS survey
(Silverman et al. 2009). As already pointed out in Fig. 8, the
distribution of the SFR in the AGN sample mirrors the one of the MIPS
selected sample (Figure 9, lower-left panel) suggesting that 
AGN activity and star formation are indeed tightly related phenomena. \\

%*****************************************
   \begin{figure*}[!ht]
 \includegraphics[width=8.7cm]{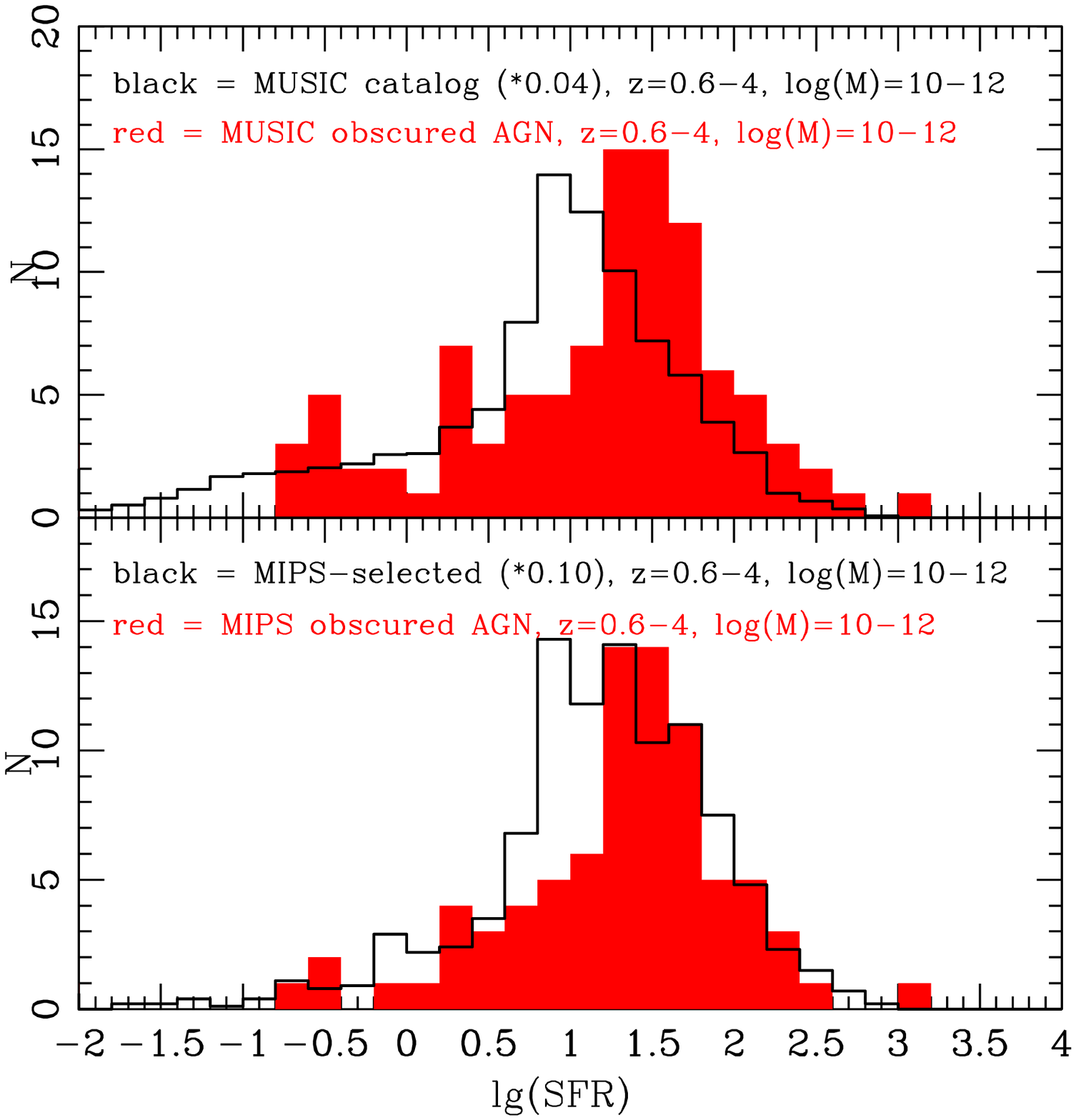}
 \includegraphics[width=8.7cm]{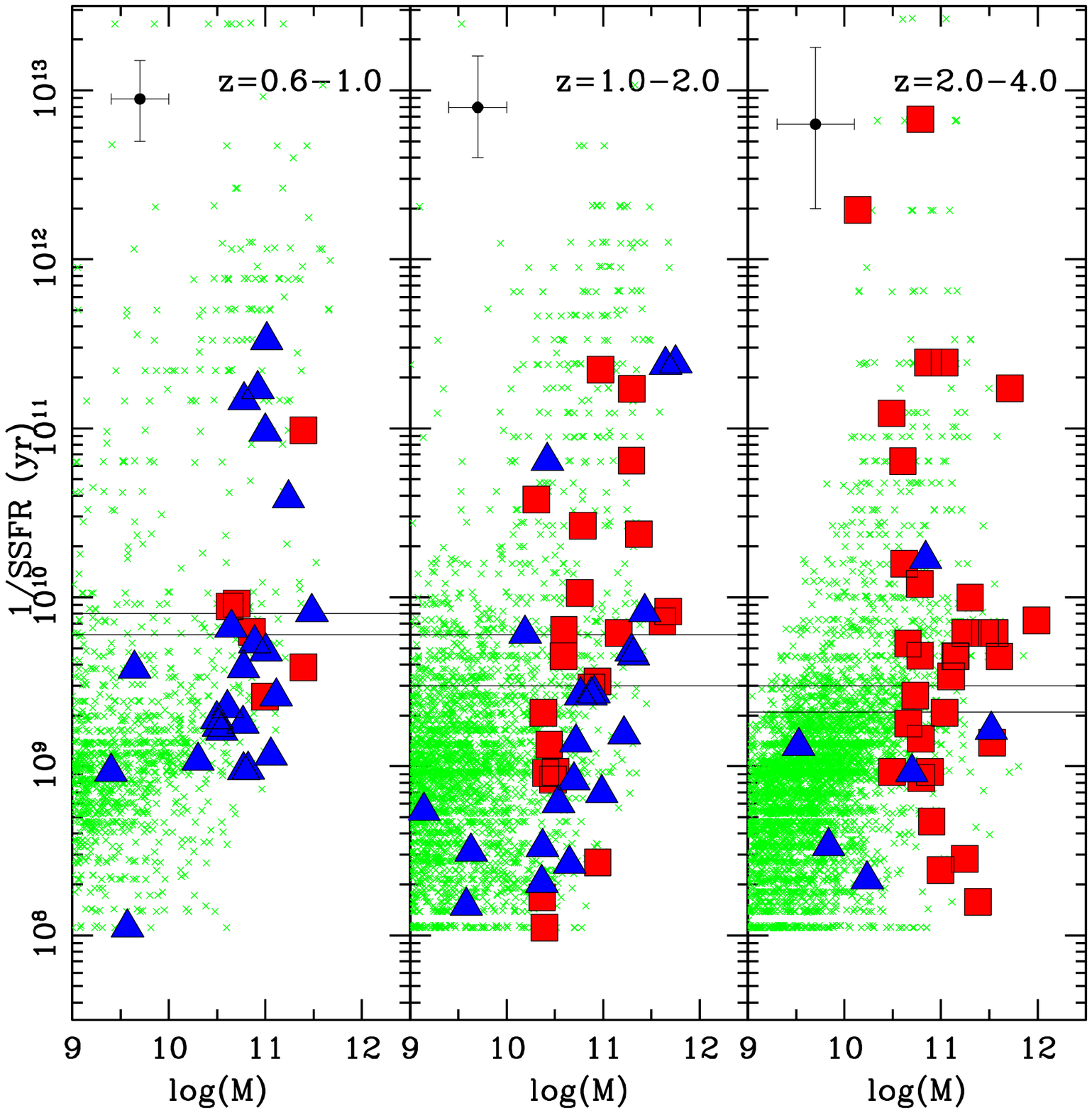}
   \caption{{\it (Left panel)}: The distribution of the SFR
for both the optically (upper) and MIPS (lower) selected sample of 
field galaxies  (black open histogram) and obscured AGN (red filled histogram) 
in the  redshift interval z=1-4 and in a specific stellar mass range 
(M$_*=10^{10}-10^{12}$  M$_\odot$).  
   {\it (Right Panel)}: inverse of the SSFR rate (defined as the inverse of 
the SFR per unit mass, M/SFR) as a function of the stellar mass of the 
galaxies in three different redshift bins for the overall field
population (green crosses) and for the obscured AGN sample
(filled symbols, color-coded on the basis of the AGN X--ray luminosity,
see Fig. 5). In each panel the representative statistical error
is also plotted as reference. 
The horizontal lines mark the age of the Universe 
at the two redshift boundaries of the chosen intervals. }
   \label{FigGam}%
    \end{figure*}

For each AGN host galaxy we could then calculate the Specific Star-Formation 
Rate (SSFR), defined as the instantaneous ratio of SFR to the total
stellar mass (SSFR=SFR/M$_*$), and its inverse, 1/SSFR, the so called 
'growth time', i.e. the time it would take a galaxy to double
its stellar mass if forming stars at the observed instantaneous rate. 
Objects with growth time lower than the age of the Universe at their
redshift (1/SSFR $<t_{Hubble}$) are actively forming stars (see, e.g. 
Fontana et al. 2009), while objects with growth time larger 
than the age of the Universe (1/SSFR $>t_{Hubble}$) can
be considered dominated by a quiescent population. 
We will name hereafter 'active host galaxies' the former, 
and 'inactive host galaxies' the latter, when referring to the
AGN hosts.\\ 
The right panel of Figure 9 shows the growth time as 
a function of the stellar mass of the galaxies in three different
redshift bins for the overall field population (green crosses) and for
the obscured AGN sample (filled symbols, color-coded on the basis of
the X--ray luminosity, see Fig.~5). The horizontal lines mark the age
of the Universe at the two redshift boundaries of the chosen intervals.
Taking the estimates of the SSFR at their face values, the majority 
($\geq50$\%) of obscured AGN in each redshift interval is hosted in 
'active host galaxies' confirming the trend outlined in Fig. 9 (left
panel). 
However, we should note that in the highest redshift bin, the SSFR 
estimates suffer also from large statistical uncertainties (see error bars 
in the upper-left part of each panel in Figure 9, right), and, 
as a consequence, the distinction 
between  'active' and 'inactive' host galaxies is rather fuzzy. Therefore, the
results for this bin should be taken only as indicative.

\subsection{AGN fraction} 

In the following we derive the AGN fraction as a function 
of the stellar mass of the underlying galaxy population. 
The AGN fraction has been computed
as the ratio of X-ray selected AGN over the overall field population
in each mass bin. 
Given that at z$>2$ we can sample only objects with 
L$_X>10^{43}$ erg s$^{-1}$, the fractions have been computed 
further imposing this threshold to the obscured AGN sample 
(i.e. considering only objects marked as squares in the figures
of the paper) in all redshifts bins 
and, therefore, they should be considered representative of this AGN 
population only\footnote{This threshold further assures that the
observed X-ray emission is mostly due to AGN activity, with a
negligible contribution from star-formation, even for the objects with
the highest observed SFR ($>1000$ M$_\odot$ yr$^{-1}$, Ranalli et
al. 2003).}\\ 
Figure 10 shows the AGN fraction as a function of M$_*$ in three
different redshift bins.  
We find that the fraction of objects hosting AGN
increases with the stellar mass of the galaxy, being 
up to 30\% for M$_*\geq 3\times10^{11}$ M$_\odot$, while it is $< 1$\% for
M$_*=10^{10}-3\times 10^{10}$ M$_\odot$.  
We should note that the MUSIC sample cannot be considered complete for 
galaxies with masses $\geq10^{10}$ M$\odot$ already at z$\sim1$. 
Indeed, different levels of 
completeness affect different kind of galaxies. The total sample is expected 
to be complete against passive galaxies with M$_{*}\geq10^{11}$ M$\odot$ 
up to z=4. In particular, the stellar mass limit of star-forming galaxies, 
which have M/L lower than passive galaxies, is lower than that of passive 
galaxies at all redshifts (Fontana et al. 2006).  
The incompleteness of 
the MUSIC sample for M$_{*}\geq10^{11}$ M$\odot$ galaxies implies that the 
plotted AGN fractions are somewhat higher than the real fractions 
at those masses, and that the actual trends of increasing AGN fraction 
as a function of the stellar mass is probably steeper than those in Fig. 10. \\
\par\noindent 
The observed trend of the AGN fraction increasing with the stellar mass
means that, for a given mass-selected sample, AGN will populate 
the high-mass tail of the parent sample mass distribution.
Qualitatively similar results are obtained by Bundy et al. 
(2008) and Yamada et al. (2009).   
This therefore also explains the position of the AGN in the left panel of Fig. 8,
i.e. among the most luminous, massive and reddest sources. 
Best et al. (2005) found a similar behaviour for the AGN fraction as a
function of the stellar mass in SDSS AGN samples, selected on the
basis of line ratios diagnostic diagrams (Baldwin, Phillips \&
Terlevich 1981; Kauffmann et al. 2003). 
In Fig. 10 we also plot the results from Best et al. (2005) 
corresponding to AGN with L([OIII]) luminosity larger than $10^{8.0}$ 
and $10^{8.5}$ L$_\odot$, as labeled.  
Using the relation between the observed L(OIII) and the X--ray luminosity 
given by Netzer et al. (2006, eq. 1), these L([OIII]) luminosities 
represent a fiducial range for local objects with X--ray luminosities 
similar to those in our sample 
(L$_X>10^{43}$ erg s$^{-1}$) when an average extinction 
E(B-V)=0.4$\pm0.2$ is taken into account (see discussion in Netzer et al. 2006).\\
It is interesting to note, however, that 
the increase with mass of the fraction of radio selected AGN
(i.e. objects in which the radio emission is not related to 
star formation processes) is much steeper than what we find for our sample 
of AGN. This is true both locally (Best et al. 2005) and up to
z$\sim1.3$ (Smolcic et al. 2009). This difference in the mass-dependence 
of the fraction of radio versus X-ray or optically selected AGN is likely
to be due to different accretion modes in these different classes
of AGN (Smolcic et al. 2009). \\

\subsection{Eddington ratio distribution function}
From our multiwavelength database we were able to estimate 
another important parameter for the characterization of 
AGN emission, namely its Eddington ratio (L/L$_{Edd}$), 
which gives the ratio at which the AGN is emitting compared 
to its expected Eddington luminosity (L$_{Edd}$). 
Indeed, the ratio between the X-ray luminosity and the host galaxy 
stellar mass is proportional to the AGN Eddington ratio,  
the proportionality factor depending on the bolometric correction 
between the X--ray and the  bolometric luminosity, and on the 
M$_{BH}$/M$_*$ ratio. We derived the Eddington ratio distribution
function for our sources as explained below.  \\
First, we calculated the ratio between the X--ray luminosity and the stellar
mass for each obscured AGN with M$_*>10^{10}$ M$_\odot$ in our sample (L$_X$/M$_*$).
Figure 11 shows the distribution of this quantity for the sources
in two different redshifts bins at z$>1$ (solid histograms in 
the upper left and bottom panels of Figure 11). 
The median (seminterquartiles) values of the distributions of logL$_X$/M$_*$ 
are similar, logL$_X$/M$_*$=32.21 (0.61) 
                           %%%% 32.87 when the Lx threshold is imposed 
and logL$_X$/M$_*$=32.73 (0.42)      
                           %%%% 32.83 when the Lx threshold is imposed  
for the two redshift bins, respectively. 
When only the sources with L$_X>10^{43}$ erg s$^{-1}$ are considered,
the median values of the logL$_X$/M$_*$ distribution are almost 
identical, logL$_X$/M$_*$=32.87 and  logL$_X$/M$_*$=32.83, respectively 
(dashed histograms). \\
In the z=1-2 redshift bin, where the uncertainties on
the SSFR estimates are reasonably small and therefore we
are able to divide the sample in 'active' and 'inactive' galaxies
(see Section 4.2), we also plotted the distribution of L$_X$/M$_*$ 
for the most luminous sources ( L$_X>10^{43}$ erg s$^{-1}$) separately 
for these two populations (upper right panel of Figure 11). 
We found that AGN in 'inactive' host galaxies tend to 
have lower log(L$_{X}$/M$_*$) values (dashed histogram in 
the upper right panel of Figure 11) than those in 'active' host galaxies 
although the statistical significance of this result is only
marginal, the probability that the two distributions are drawn from the same 
parent population being 5\%, according to a Kolmogorov-Smirnov statistics). \\
Then, we assumed the local relation between M$_{BH}$ and M$_*$ 
(e.g Marconi \& Hunt 2003,  M$_{BH}\sim10^{-3}\times M_*$) to convert the stellar mass in a BH mass
and calculate L$_{Edd}$, and we applied a bolometric correction of 20 
(e.g. Marconi et al. 2004) to derive the bolometric luminosity
from the X--ray luminosity. 
With these assumptions, the median values of L$_X$/M$_*$ correspond 
to L/L$_{Edd}\sim$0.02-0.08 (L/L$_{Edd}\sim$0.1 for the logLx$>43$ 
subsample).
We note that, given the systematics uncertainties affecting the derivation 
of the stellar masses (see discussion in 2.4), and the fact that {\it total} stellar
masses are derived and not {\it bulge} masses, the derived BH massses are most
likely upper limits, this translating in conservative (lower-limits) estimate of
the  L/L$_{Edd}$.  
On the other hand, recently, it has been suggested that the
average M$_{BH}$-host galaxy mass ratio at z=1-2 is higher by a factor $\sim2$
than the local value (McLure et al. 2006, Peng et al. 2006, Merloni 
et al. 2010). Should this be the case, our estimates of L/LEdd 
should be reduced by a factor of $\sim 2$ at z=1-2 and possibly more
in the higher redshift bin if the average  M$_{BH}$-M$_*$ ratio
further increases to higher redshift.

\par\noindent

   \begin{figure}[!t]
 \includegraphics[width=8cm]{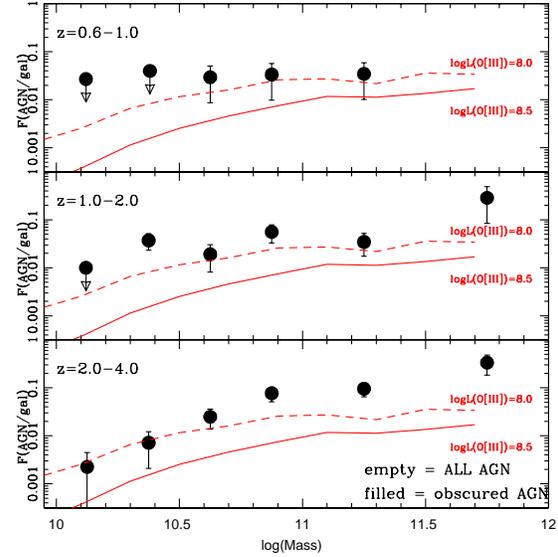}
   \caption{Fraction of obscured AGN with L$_{X}>10^{43}$
erg s$^{-1}$ as a function
of the stellar mass in three different redshift bins (see
Fig. 8 for the definition). Filled circles refer to the ``obscured'' AGN sample.
The dashed (continuous) lines represent the fraction of AGN with L(OIII)$>10^{8.0} (10^{8.5})$ L$_\odot$
in an optically selected sample in the local Universe (z$<0.2$) from the SDSS
(Best et al. 2005). See text for  details.}

   \label{FigGam}%
    \end{figure}

   \begin{figure}[!t]
 \includegraphics[width=8cm]{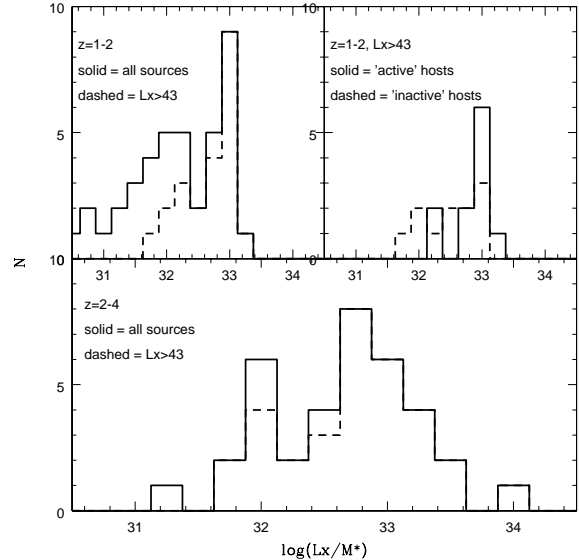}
   \caption{The distribution of the ratio between  the X--ray luminosity and the stellar
mass for the obscured AGN 
in our sample (empty histograms), 
in two different redshifts bins as labeled. }
   \label{FigGam}%
    \end{figure}

\section{Discussions and Conclusions}

We presented a new catalog of the counterparts of the 179
extragalactic X--ray sources detected in the 1Ms Chandra observation
of the MUSIC/CDFS/GOODS field and an extensive analysis of the host
galaxies properties of obscured AGN. \\ 
We quantified the bias in the determination of the counterparts of
X--ray selected sources when the match is limited to optical catalogs
only, with respect to the combined use of optical and near infrared
(deep K--band and IRAC) data.  We estimate that the fraction of
misidentified X--ray sources previously reported in the literature is
of the order of $\sim 6\%$, and rises up to $\sim14$\% when optically
faint ($z>24$) sources are considered (see Figure 1); 
the use of an optically-based catalog
biases the identification against the most extreme, obscured sources
therefore preventing the exact knowledge of the multiwavelength
properties of the X-ray counterparts. \\
\par\noindent 
In order to study the host-galaxies of obscured AGN, 
we defined a sample of 116 ``bona fide'' obscured AGN, by selecting sources
without broad lines in the optical spectra and with small 
optical nuclear emission with respect to the host galaxy optical emission
(see Figure 3).
From eyeball inspection of the host galaxy morphology, we found a
variety of cases, and a disturbed morphology (due to
activity/merging/star formation) in more than half of the sub-sample
for which a morphological classification could be made 
(see Figure 4).\\
\par\noindent
%%%%%%%%%%%%%%%%%%%%%%%%%%%%%%%%%%%%%%%%%%%%%%%%%%%%%%%%%%%%%%%%%%%%%%%%%%%%%%%%%%%%%%%%%%%%%%%%%%%%%%%%%%%%%
We investigated the optical to infrared colors of these 
obscured AGN. 
The most striking result is that half of the X--ray 
selected obscured AGN in the redshift range z=1.5-4.0 
have a 8.0$\mu$m to 4.5$\mu$m flux ratio $<2$ and 
according to Pope et al. (2008) would have been classified 
as ``star-formation'' dominated objects (see Figure 6); 
50\% of them in addition have 
24.0$\mu$m to 8.0$\mu$m flux ratio $<5$, where the original 
Pope et al. (2008) diagram is almost empty. 
Moreover, previous analysis based on Chandra stacking analysis of sources 
with high MIR/O (Daddi et al. 2007a, Fiore et al. 2008, 2009)
claimed a large contribution ($\geq80$\%) of heavily obscured (Compton Thick) 
sources among the stacked population (but see also discussion in
Donley et al. 2008). 
This suggests that 1) the accretion activity in high-redshift sources
is unambiguously revealed thanks to the presence of a strong X--ray 
emission, and 2) the star-formation region as defined by Pope et al. (2008)
contains not only objects 
in which the bolometric luminosity is dominated by the star-formation 
processes, but also a not-negligible number of objects hosting
candidate obscured/Compton Thick  AGN.  
In conclusion, our obscured AGN show Spitzer colors 
consistent with both an AGN dominated continuum and a starburst 
dominated continuum in the MIR. \\
\par\noindent
%%%%%%%%%%%%%%%%%%%%%%%%%%%%%%%%%%%%%%%%%%%%%%%%%%%%%%%%%%%%%%%%%%%%%%%%%%%%%%%%%%%%%%%%%%%%%%%%%%%%%%%%%%%%%
We found that the hosts of obscured AGN are redder in (U-V) rest frame 
than the overall galaxy population at the same redshift: in particular, 
obscured AGN mainly
populate the red sequence and the green valley in the color-magnitude
plots (Figure 7, left panel), 
in agreement with the results of Nandra et al. (2007), and
Silverman et al. (2007).
For the MUSIC sample the U-V galaxy colors are
strongly correlated with the K band absolute magnitude
(Figure 8, left panel), and therefore
with the galaxy stellar mass, with the most massive systems having a
redder color.  The hosts of the obscured
AGN are therefore found in the red-massive tail of the distribution of 
optically selected galaxies in all three redshift bins considered.  
AGN feedback is often invoked as one
of the main responsible for the observed galaxy colors (Nandra et al. 2007,
Hasinger 2008). 
However, it is well known that the main ingredient for nuclear
activity is the presence of a SMBH in galaxy
nuclei, and that SMBHs are found nearly exclusively in
massive galaxies (e.g. Magorrian et al. 1998). 
Therefore, it is not truly surprising to find AGN hosted 
in massive galaxies and the simple presence of AGN in massive red galaxies 
is not enough to argue for a significant feedback effect on the 
observed colors, because of the strong color-mass correlation. 
Were AGN feedback responsible for the observed red colors, since
galaxy colors are strongly correlated with the galaxy mass and AGN are
found preferably in massive galaxies, then AGN feedback
should be also considered as one of the main players in the building 
of the galaxy mass-color correlation. \\
\par\noindent
%%%%%%%%%%%%%%%%%%%%%%%%%%%%%%%%%%%%%%%%%%%%%%%%%%%%%%%%%%%%%%%%%%%%%%%%%%%%%%%%%%%%%%%%%%%%%%%%%%%%%%%%%%%%%
We found that about 2/3 of the obscured AGN
hosts at all redshifts show substantial ($>10$ M$_\odot$ yr$^{-1}$)
star formation activity (Figure 9 left panel) and about half 
live in galaxies which are still actively forming 
stars with respect to their mass (Figure 9, right panel). 
For these sources, the observed red colors are
likely due to dust extinction rather than evolved stellar population. 
We then conclude that a significant fraction of obscured AGN 
live in massive, dusty star-forming galaxies with red optical colors. 
This result is in qualitative agreement with the morphological analysis. 
Higher luminosity X-ray selected AGN are not systematically found in objects 
with the highest SSFR (see Fig. 9 right panel), 
in agreement with Alonso-Herrero et al. (2008). \\
\par\noindent 
%%%%%%%%%%%%%%%%%%%%%%%%%%%%%%%%%%%%%%%%%%%%%%%%%%%%%%%%%%%%%%%%%%%%%%%%%%%%%%%%%%%%%%%%%%%%%%%%%%%%%%%%%%%%%
We compared the number of obscured AGN and of all X--ray selected AGN
to the number of field galaxies in broad bins of galaxy stellar mass
(M$_*=10^{10}-10^{12}$ M$_\odot$) and redshifts (z=0.6-1, z=1-2,
z=2-4). We find that the AGN fraction increases with the host
galaxy stellar mass, from $\sim$1\% at M$_*\sim 10^{10}$ M$_\odot$ to
$\sim$30\% at M$_*\sim 3\times10^{11}$ M$_\odot$ (see also Yamada et al. 2009),
and the actual trend 
of increasing AGN fraction  as a function of the stellar mass is probably 
steeper given the uncompleteness of the MUSIC sample at M$_{*}<10^{11}$ 
M$\odot$ (see Section 4.4). The uncertainties on the stellar
mass estimate from the SED fitting have the effect of 
shifting at lower masses (of $\sim 0.25$ dex) the datapoints, leaving
the total fraction and the slope unchanged. \\
We compared this trend with
that observed in the local Universe (Best et al. 2005) for AGN with 
luminosity above similar thresholds. While the observed trend is the same,
in all the investigated redshift bins the AGN fraction is 
higher than that observed in the local Universe, and it could likely be 
even higher.  
In fact, we are comparing here AGN selected with two different methods: forbidden
emission line luminosity (SDSS) and X-ray emission (GOODS). The latter
sample does not contain most Compton thick AGN.  
On the other hand, Compton thick AGN may well be present in [OIII] selected
AGN samples. 
The fraction of Compton thick AGN not directly
detected in deep Chandra surveys is estimated between 40\% and 100\%
of the X-ray selected AGN, using infrared selection or other
techniques (see Donley et al. 2008, Fiore et al. 2009 and references therein).
Therefore, under the simplest assumption that this Compton Thick AGN fraction is constant
with the galaxy mass, the discrepancy observed in Figure 10 can increase by up
of a factor of 2. \\
The fraction of active galaxies to the total galaxy population is
proportional to the AGN duty cycle. Our results would thus suggest
higher AGN duty cycles at z=2-4 than at z=0, in agreement with 
expectations from most recent semi-analytic models (e.g. Menci et al. 2008), 
in which at higher redshift the AGN activity is present in a 
larger number of galaxies than locally.  \\
\par\noindent
%%%%%%%%%%%%%%%%%%%%%%%%%%%%%%%%%%%%%%%%%%%%%%%%%%%%%%%%%%%%%%%%%%%%%%%%%%%%%%%%%%%%%%%%%%%%%%%%%%%%%%%%%%%%%
The fact that the most luminous obscured AGN are found in the most massive galaxies
at all investigated redshifts may suggest that the L/L$_{Edd}$ of the obscured AGN 
is similar, particularly in the case of the most luminous sources (logLx$>43$ erg s$^{-1}$),
for which the threshold in luminosity introduces a bias against the sources 
accreting at lower rates in the lowest redshift bin. Assuming the local Magorrian 
relation between  M$_{BH}$ and M$_*$ (e.g Marconi \& Hunt 2003) and a bolometric 
correction of 20 (e.g. Marconi et al. 2004) the median values of L$_X$/M$_*$=32.83  
and L$_X$/M$_*$=32.87 correspond to L/L$_{Edd}\sim0.1$. 
Although suffering from large uncertainties associated with the stellar mass and BH mass 
estimates (see Section 4.4), this value can be considered as 
representative of the accretion state of the most luminous, 
obscured AGN in the present sample.
Similar results are obtained from a comprehensive analysis of host galaxies properties of 
Chandra Deep Field North X--ray sources sources at z=2-4 (Yamada et al. 2009) 
and are also typical of unobscured Type 1 AGN at z$>1$ (Merloni et al. 2010, Trump et al. 2009). \\
It is instructive to compare these findings with similar estimates 
in the local Universe.  Kauffmann \& Heckman (2009) using SDSS data
found an average L/L$_{Edd}$ value of $\sim 0.01$ and a log normal distribution for this parameter, 
for AGN hosted in galaxies with significant on-going star formation, 
while AGN in inactive host galaxies follow a power-law distribution.  
As discussed in section 4.4, and shown in the upper right panel of figure 11,
we also find marginal evidence of different accretion rates distributions 
for the populations of AGN in 'active' and 'inactive' host galaxies. \\

\section{Summary}

Taking advantage of the 1Ms CDFS Chandra observations and the associated 
deep multiwavelength follow-up, we have studied the host galaxies properties 
of obscured AGN at z$>1$.  \\
The most important results of our analysis can be summarized as follows:

\begin{itemize} 

\item X--ray selected (L$_X\gs10^{42}$ erg s$^{-1}$)  
AGN show Spitzer colors consistent with both AGN and starburst dominated
infrared continuum and only the combination of the
two selections (X--ray + Mid-infrared) can fully characterize the growth of SMBH at z$>1$; \\
 
\item the host galaxies of X--ray selected obscured AGN are all massive
and in 50\% of the cases are also actively forming stars. Their red
colors are due to dust extinction; \\

\item the X--ray selected AGN fraction increases with the stellar mass up to a value
of $\sim30$\% at z$>1$ and M$_*>3\times10^{11}$ M$\odot$, a fraction significantly
higher than what is observed in the local Universe ($\sim 2$\%) for AGN of similar 
luminosities; 

\item the median L/L$_{Edd}$ value is between 2\% and 10\% depending on
the assumed M$_{BH}$/M$_{*}$ ratio, the X--ray luminosity threshold and 
the host galaxies properties of the obscured AGNs. \\

\end{itemize} 
   
\footnotesize{
\begin{table*}
\caption{Counterparts to CDFS X-ray sources: positions, fluxes, redshifts and luminosities}\label{cptable}
%\centering
\begin{tabular}{rrrccrrrrrr} 
\hline\hline             
AID & XID & OID & RA & DEC & dist & FLAG & S$_{2-10}$ keV & S$_{0.5-2}$ keV & redshift & log(L$_{X}$)\\ 
    &     &     & J2000 & J2000 & arcsec & & 10$^{-15}$ cgs & 10$^{-15}$  cgs &  & erg s$^{-1}$ \\
\hline
   51 &   108a &    13272 &   53.024029 &  -27.746366 &  0.32  & 0$^m$  &   1.13 &   0.41                 & 1.608    (3) & 	         43.33 \\
   53 &   560* &    12222 &   53.026127 &  -27.760183 &  0.22  &  0 &  $<0.51$ &   0.15   		  & 0.669     (1) & 		 41.63 \\
   67 &   221 &    70103 &   53.036953 &  -27.740129 &  0.34  &  0$^{k,m}$ &  $<0.52$ &   0.15   	  & 2.43 (0.7-4.5)	 & 	 43.13 \\
   71 &    61 &    70051 &   53.043751 &  -27.719135 &  0.12  &  0 &  12.28 &   5.68    		  & 3.59  (3.4-4.3)	 & 	 45.04 \\
   73 &   536 &    15717 &   53.044842 &  -27.709606 &  0.34  &  0 &  $<0.87$ &   0.11   		  & 0.418    (4) & 		 41.26 \\
   75 &   185 &    14349 &   53.045467 &  -27.728628 &  1.66  &  0 &  $<0.61$ &   0.09   		  & 0.998    (3) & 		 42.55 \\
   76 &    60 &    13849 &   53.045471 &  -27.737484 &  0.16  &  0 &  11.89 &   6.66    		  & 1.615     (1) & 		 44.23 \\
   77 &    80 &     8168 &   53.045666 &  -27.815575 &  0.03  &  0 &   1.64 &   0.93    		  & 1.39   (1.2-3.0)	 & 	 43.21 \\
   80 &   535 &    10619 &   53.047535 &  -27.780546 &  0.21  &  0 &   1.28 &   0.26    		  & 0.575     (1) & 		 42.23 \\
   82 &    58 &    11180 &   53.049049 &  -27.774496 &  0.02  &  0$^m$  &   2.18 &   0.79   		  & 2.21  (1.6-4.0)	 & 	 43.81 \\
   83 &   534 &    12281 &   53.050819 &  -27.758352 &  0.73  &  0 &  $<0.55$ &   0.07   		  & 0.676	   (2) & 	 42.27 \\
   84 &   149 &    11285 &   53.050934 &  -27.772402 &  0.42  &  0 &   1.19 &   0.17    		  & 1.033	   (2) & 	 42.79 \\
   88 &    56a &    15626 &   53.055141 &  -27.711374 &  0.09  &  0 &  17.50 &   2.81   		  & 0.605     (1) & 		  43.4 \\
   90 &   600 &    12434 &   53.057587 &  -27.757130 &  0.14  &  0 &   0.66 &  $<0.06$   		  & 1.327	   (2) & 	 43.13 \\
   91 &   266 &    15382 &   53.057747 &  -27.713591 &  0.73  &  0 &   1.01 &  $<0.08$   		  & 0.735     (1) & 		 42.81 \\
   93 &   583 &    70323 &   53.057854 &  -27.833355 &  0.78  &  0$^{k,m}$ &  $<0.48$ &   0.19   	   & 2.24  (1.2-2.3)	 & 	  43.6 \\
   94 &    55 &     5434 &   53.058353 &  -27.850199 &  0.12  &  0 &   7.78 &   0.83    		  & 0.122     (1) & 		 41.51 \\
   95 &   532 &    16040 &   53.058666 &  -27.708439 &  0.17  &  0 &  $<0.85$ &   0.16   		  & 2.026    (4) & 		 42.83 \\
   96 &   531 &     5281 &   53.060116 &  -27.853006 &  0.18   & 0 &   0.97 &   0.07    		  & 1.544     (1) & 		 43.46 \\
   99 &   593 &    14104 &   53.061623 &  -27.734022 &  1.84   & 0$^m$ &  $<0.59$ &  $<0.10$    	   & 1.67  (1.3-2.25)	   & 	 42.44 \\
  100 &    82a &     5440 &   53.061905 &  -27.851044 &  0.17   & 0$^m$ &   1.46 &   0.27   		    & 1.21  (0.8-2.6)	   & 	 43.05 \\
  102 &    83 &    16094 &   53.062420 &  -27.706909 &  0.29   & 0 &   3.29 &   1.04    		   & 2.494    (4) & 		 43.56 \\
  103 &    53 &     4878 &   53.062420 &  -27.857515 &  0.18   & 0 &   4.00 &   1.93    		   & 0.675     (1) & 		 42.86 \\
  106 &   552 &     2654 &   53.065826 &  -27.890200 &  0.35   & 0 &  $<0.40$ &   0.28   		   & 0.674     (1) & 		 41.91 \\
  111 &   218b &     4384 &   53.068401 &  -27.866474 &  0.33   & 0 &  $<0.56$ &   0.29   		   & 1.48   (1.7-2.6)	  & 	 42.72 \\
  112 &   564 &    10944 &   53.069603 &  -27.777264 &  0.09   & 0$^m$ &   0.31 &   0.12   		      & 2.61  (0.5-3.4)      &  43.14  \\
  113 &   173 &    14774 &   53.069737 &  -27.724312 &  0.37   & 0 &  $<0.58$ &   0.10   		    & 0.524	(1) &  	 41.56	       \\
  114 &   635a &     6606 &   53.069878 &  -27.835489 &  1.73   & 0 &  $<0.32$ &   0.04   		    & 0.729	(1) &  	 41.49	       \\
  115 &   575 &     7393 &   53.071068 &  -27.822748 &  0.43   & 0 &   0.47 &   0.06    		   & 0.340     (1) & 		 41.26 \\
  116 &   205a &    17038 &   53.071293 &  -27.693579 &  0.29   & 0 &   1.29 &   0.17   		   & 2.22    (1.5-2.7)    & 	 43.62 \\
  117 &    52 &    15207 &   53.071426 &  -27.717581 &  0.02   & 0 &   7.89 &   4.85    		   & 0.569     (1) & 		 42.98 \\
  118 &    51 &     3920 &   53.071545 &  -27.872448 &  0.13   & 0 &  14.16 &   0.62    		   & 1.097     (1) & 		 44.02 \\
  121 &  566  &     9792 &   53.075058 &  -27.788486 &  0.83   & 0 &  $<0.41$ &   0.30   		   & 0.734	    (2) & 	 42.02 \\
  122 &    87 &     3533 &   53.076000 &  -27.878160 &  0.12   & 0 &   0.76 &   0.39    		   & 2.801     (1) & 		 43.58 \\
  123 &   153 &     5551 &   53.076405 &  -27.848669 &  0.05   & 0 &   5.06 &   0.19    		   & 1.536     (1) & 		 44.17 \\
  124 &   526a &    13866 &   53.077923 &  -27.736874 &  1.06   & 0 &   0.66 &   0.09   		   & 0.958     (1) & 		 42.46 \\
  125 &   263b &     4835 &   53.078468 &  -27.859856 &  0.73   & 0$^m$ &   1.11 &   0.08   		   & 3.660     (1) & 		 44.13 \\
  126 &    50 &     9279 &   53.079144 &  -27.798740 &  0.10   & 0 &   2.21 &   0.37    		   & 0.670     (1) & 		 42.61 \\
  129 &   525 &    17067 &   53.082535 &  -27.689650 &  0.36   & 0 &  $<0.97$ &   0.25   		   & 0.229     (1) & 		 40.78 \\
  130 &   524 &    70052 &   53.083061 &  -27.711782 &  0.85   & 0$^m$ &   1.23 &   0.22   		    & 1.50  (0.0-5.5)	   & 	 43.19 \\
  131 &   253 &    13250 &   53.083542 &  -27.746447 &  0.32   & 0 &   2.99 &   0.13    		   & 0.481     (1) &		 42.47 \\
  133 &   523 &    70044 &   53.084816 &  -27.707909 &  0.07   & 0$^{k,m}$ &   0.65 &   0.13    	    & $>3.50$	     & 	 $>43.76$	       \\
  134 &   151 &     9723 &   53.085323 &  -27.792313 &  0.14   & 0 &   4.34 &   0.07    		   & 0.604	    (2) & 	 42.98 \\
  136 &   522a* &      489 &   53.089176 &  -27.930479 &  0.43   & 0 &  $<0.97$ &   0.47   		   & 2.611    (4) & 		 43.73 \\
  137 &   612b &    15948 &   53.089264 &  -27.708660 &  0.87   & 0 &   1.24 &  $<0.08$   		   & 0.736     (1) & 		  42.8 \\
  139 &   602 &    10384 &   53.091606 &  -27.782188 &  0.60   & 0 &   0.38 &  $<0.06$   		   & 0.668     (1) & 		 42.34 \\
  141 &   572 &    70249 &   53.092484 &  -27.803045 &  0.71   & 0$^m$ &  $<0.50$ &   0.07   		   & 2.75   (1.5-3.8)	  & 	 43.37 \\
  143 &   561a &    12146 &   53.093548 &  -27.762192 &  0.13   & 0$^m$ &  $<0.36$ &   0.10   		   & 2.14   (1.1-3.8)	  & 	 42.64 \\
  145 &   145 &    11723 &   53.093922 &  -27.767733 &  0.07   & 0 &   3.84 &   0.38    		   & 1.727     (6) & 		 43.87 \\
  146 &   188 &     7031 &   53.093948 &  -27.830509 &  0.16   & 0 &   0.59 &   0.10    		   & 0.734     (1) & 		 42.15 \\
  149 &   521 &     3781 &   53.094841 &  -27.873321 &  0.50   & 0 &   0.81 &   0.26    		   & 0.131     (1) & 		 40.56 \\
  151 &   204 &    11854 &   53.096493 &  -27.765190 &  0.26   & 0 &   0.40 &   0.07    		   & 1.223	    (2) & 	 42.51 \\
  152 &   627 &     8029 &   53.097259 &  -27.814608 &  0.52   & 0 &  $<0.30$ &   0.07   		   & 0.248     (1) & 		 40.94 \\
  154 &   241 &    15419 &   53.100811 &  -27.715981 &  0.10   & 0 &  $<0.65$ &   0.26   		   & 2.303    (4) & 		 43.12 \\
  155 &    49 &    17077 &   53.101070 &  -27.690670 &  0.35   & 0 &   3.76 &   2.01    		   & 0.534     (1) & 		 42.59 \\
  157 &   598 &     1780 &   53.102806 &  -27.903250 &  0.10   & 0 &  $<0.62$ &  $<0.09$   		   & 0.617     (1) & 		 42.72 \\
  158 &   565 &    10250 &   53.103580 &  -27.785107 &  0.35   & 0 &  $<0.38$ &   0.10   		   & 0.363	    (2) & 	 40.93 \\
  159 &    48 &      371 &   53.103531 &  -27.933342 &  0.13   & 0 &   4.83 &   1.03    		   & 1.049    (3) & 		  43.4 \\
  160 &   606a &     6640 &   53.103993 &  -27.835571 &  0.19   & 0$^m$ &   1.19 &   0.07   		   & 1.037     (1) & 		 42.88 \\
  162 &   260 &     5832 &   53.104610 &  -27.845348 &  0.24   & 0 &   1.09 &   0.09    		   & 1.043     (1) & 		 42.93 \\
  163 &    46 &    16224 &   53.104858 &  -27.705219 &  0.05   & 0 &   4.24 &   2.99    		   & 1.617     (1) & 		 43.78 \\
  164 &   150 &     1192 &   53.104832 &  -27.913927 &  0.55   & 0 &   2.62 &  $<0.09$   		   & 1.090     (1) & 		 43.33 \\
  166 &    45 &    15260 &   53.107006 &  -27.718241 &  0.14   & 0$^m$ &   5.04 &   1.22   		   & 2.291     (1) & 		 44.22 \\
  167 &   233 &     7194 &   53.107262 &  -27.826765 &  0.25   & 0 &  $<0.66$ &   0.09          	   & 0.577     (1) & 		 42.14 \\
\hline
\end{tabular}
\end{table*}}
\footnotesize{
\begin{table*}
%\caption{IRAC counterparts to CDFS X-ray sources - positions}\label{cptable}
%\centering
\begin{tabular}{rrrccrrrrrr} 
\hline\hline             
AID & XID & OID & RA & DEC & dist & FLAG & S$_{2-10}$ keV & S$_{0.5-2}$ keV & redshift & log(L$_{X}$)\\ 
    &     &     & J2000 & J2000 & arcsec & & 10$^{-15}$ cgs & 10$^{-15}$  cgs &  & erg s$^{-1}$\\
\hline
  170 &   589 &    70385 &   53.107422 &  -27.855698 &  0.66   & 0$^{k,m}$ &  $<0.38$ &   0.10             & 3.98 (1.25-4.8)&                 43.24 \\
  171 &   519 &      978 &   53.107746 &  -27.918447 &  1.09   & 0 &  $<0.80$ &   0.20   		      & 1.034	  (1) & 	       42.41\\
  172 &    81 &    12747 &   53.108109 &  -27.753985 &  0.21   & 0$^m$ &   1.12 &   0.52   		     &     2.98   (2.0-3.7)	&     43.82 \\
  175 &   518 &    70168 &   53.111542 &  -27.767839 &  0.26   & 0$^m$ &   0.63 &   0.10   		     &      $>0.40$	   & 	       $>41.55$\\
  176 &    43 &    16792 &   53.111519 &  -27.695988 &  0.06   & 0 &   3.57 &   0.69    		   	& 0.734     (1) & 	       42.91\\
  177 &    42a &    17452 &   53.112530 &  -27.684723 &  0.15   & 0  &  84.95 &  53.27   		      & 0.734	  (1) & 	       44.27\\
  179 &    41 &    16831 &   53.115097 &  -27.695806 &  0.19   & 0 &   8.81 &   0.55    		   	& 0.668     (1) & 	       43.24\\
  180 &   623a &    10464 &   53.118420 &  -27.783386 &  0.70   & 0 &  $<0.20$ &   0.08   		   	 & 2.14   (1.8-2.6)	&     42.62 \\
  181 &   224 &    11248 &   53.119724 &  -27.772304 &  0.10   & 0 &   0.27 &   0.26    		   	 & 0.735	  (2) &       41.77 \\
  182 &   103 &    13977 &   53.120071 &  -27.732124 &  0.25   & 0 &   1.07 &   0.72    		   	  & 0.215     (1) & 	       41.15\\
  187 &   599 &     2637 &   53.124100 &  -27.891216 &  0.76   & 0$^m$ &   0.67 &  $<0.08$   		   	 & 2.59  (2.0-4.0)	&     43.62 \\
  188 &   202 &     5390 &   53.124371 &  -27.851633 &  0.16   & 0 &   2.81 &   0.35    		   	 & 3.700     (1) & 	       44.55\\
  189 &    95 &    13855 &   53.124493 &  -27.740086 &  0.69   & 0 &  $<0.50$ &   0.81   		   	 & 0.076	  (2) &       40.27 \\
  191 &    39 &    12354 &   53.124916 &  -27.758301 &  0.08   & 0 &  15.57 &   8.44    		   	 & 1.218     (1) & 	       44.06\\
  192 &   116 &    13853 &   53.124950 &  -27.734684 &  0.90   & 0 &   0.89 &   0.61    		   	 & 0.076	  (2) &        40.1   \\
  193 &    78 &    12489 &   53.125252 &  -27.756536 &  0.04   & 0 &   2.75 &   1.94    		   	 & 0.960     (1) & 	       43.06  \\
  195 &    38 &    12880 &   53.125900 &  -27.751274 &  0.11   & 0 &   9.04 &   6.23    		   	 & 0.738     (1) & 	        43.3  \\
  196 &   516 &     9791 &   53.130642 &  -27.790258 &  0.19   & 0 &  $<0.42$ &   0.08   		   	  & 0.665     (1) & 	       41.91  \\
  197 &   563 &    11272 &   53.131081 &  -27.773107 &  0.23   & 0 &  $<0.37$ &   0.07   		   	 & 2.223     (1) & 	       42.64  \\
  198 &   574 &     8217 &   53.131554 &  -27.814941 &  0.47   & 0 &  $<0.34$ &   0.09   		   	 & 1.879     (6)& 	       42.56  \\
  200 &    37* &    16661 &   53.133640 &  -27.698654 &  0.16   & 0 &   2.20 &   0.64   		   	 & 1.393     (3) & 	       43.34  \\
  201 &   515 &    70195 &   53.134022 &  -27.781019 &  0.27   & 0$^m$ &   1.23 &   0.11   		   	 & 2.98   (1.9-4.6)	&      43.9   \\
  202 &   220a &    70404 &   53.136291 &  -27.864195 &  0.57   & 0 &   0.75 &   0.35   		   	   & 1.42  (1.3-1.6	 &     42.9   \\
  204 &    36a &    11970 &   53.137573 &  -27.763176 &  0.06   & 0 &   3.09 &   1.14   		   	 & 0.953    (3) & 	       43.11  \\
  205 &   217a &    70429 &   53.137924 &  -27.868242 &  0.57   & 0$^{k,m,r}$  &  $<0.50$ &   0.19   	   	 & $>3.0$      & 	       $>43.68$  \\
  206 &   265 &    15927 &   53.138687 &  -27.709997 &  0.22   & 0$^m$ &   3.24 &   0.25   		   	 & 1.213      (6) & 	       43.45  \\
  207 &   514* &    14880 &   53.139412 &  -27.720207 &  0.34   & 0 &   0.93 &   0.20   		   	& 0.103     (5) & 	        40.4  \\
  208 &    86a &    12616 &   53.140991 &  -27.755701 &  0.20   & 0 &  $<0.48$ &   0.11   		   	& 1.894    (3) & 	       42.75  \\
  209 &   513a &    70289 &   53.141506 &  -27.816603 &  0.92   & 0$^m$ &   0.64 &  $<0.06$   		   	& 2.66    (1.75-4.3)	&     43.58   \\
  210 &   579b &     7316 &   53.141880 &  -27.827168 &  0.00   & 0 &  $<0.38$ &   0.08   		   	& 0.832    (3) & 	       41.59  \\
  212 &   512 &    14262 &   53.143093 &  -27.730581 &  0.64   & 0 &   0.72 &   0.20    		   	 & 0.668     (1) & 	       42.11  \\
  214 &    34a &      921 &   53.145634 &  -27.919773 &  0.23   & 0 &   4.12 &   1.66   		   	& 0.839     (1) & 	        43.1  \\
  216 &   171 &    13946 &   53.146271 &  -27.736269 &  0.07   & 0 &  $<0.49$ &   0.14   		   	& 0.839 	 (2) &        41.82   \\
  218 &   148 &     2818 &   53.146698 &  -27.888338 &  0.36   & 0$^m$ &   2.68 &   0.44   		   	& 2.06  (1.2-2.75)	&     43.85   \\
  219 &   901 &     7814 &   53.148827 &  -27.821100 &  0.06   & 0 &   0.60 &   0.06    		   	& 2.578     (1) & 	       43.43  \\
  221 &   100 &     8231 &   53.149887 &  -27.814003 &  0.16   & 0 &   0.38 &   0.23    		   	& 1.309     (1) & 	       42.52  \\
  222 &   239a &     4993 &   53.150711 &  -27.857361 &  0.15   & 0 &  $<0.39$ &   0.21   		   	& 1.613    (3) & 	       42.75  \\
  223 &   609 &     5988 &   53.150658 &  -27.843628 &  0.20   & 0 &   0.80 &  $<0.10$   		   	& 1.608    (3) & 	       43.53  \\
  224 &   577a &     7347 &   53.150726 &  -27.825510 &  0.58   & 0 &  $<0.34$ &   0.13   		   	& 0.547     (1) & 	       41.35  \\
  226 &   511c &    11049 &   53.152779 &  -27.775293 &  0.40   & 0 &  $<0.46$ &   0.13   		   	& 0.999 	 (2) &        42.05   \\
  227 &    33 &    13964 &   53.152977 &  -27.735123 &  0.16   & 0 &   9.98 &   4.40    		   	& 0.665 	 (2) &        43.24   \\
  230 &    31 &     4119 &   53.157345 &  -27.870085 &  0.13   & 0 &  10.78 &   6.45    		   	& 1.603     (1) & 	       44.18  \\
  231 &   555 &     3010 &   53.158051 &  -27.885572 &  0.24   & 0$^m$ &  $<0.46$ &   0.09   		   	& 2.73   (2.3-3.1)     &      42.86   \\
  233 &    79 &    11237 &   53.158451 &  -27.773975 &  0.11   & 0$^m$ &   1.85 &   0.95   		   	& $>0.50$	 & 	       $>42.22$  \\
  235 &   210 &      461 &   53.159641 &  -27.931435 &  0.47   & 0$^m$ &  $<0.94$ &   0.37   		   	 & 1.61  (1.2-2.25)	 &    43.12   \\
  236 &   567a &     9696 &   53.161564 &  -27.792257 &  0.19   & 0 &  $<0.35$ &   0.09   		   	& 0.458     (5) & 	          41  \\
  237 &   510 &     5099 &   53.161503 &  -27.856007 &  0.28   & 0$^m$ &   0.82 &   0.13   		   	& 2.97  (2.8-3.7)      &      43.72   \\
  238 &   582b &     6757 &   53.161751 &  -27.832291 &  0.12   & 0 &  $<0.33$ &   0.07   		   	& 0.242    (3) & 	       40.92  \\
  241 &   201a &    70116 &   53.162647 &  -27.744331 &  0.59   & 0 &   2.18 &   0.52   		   	& 2.52   (1.5-4.0)     &      43.96   \\
  242 &    28 &    11701 &   53.162861 &  -27.767160 &  0.12   & 0 &   3.24 &   0.90    		   	  & 1.216     (1) & 	       43.38  \\
  243 &   605 &     8625 &   53.163220 &  -27.808992 &  0.32   & 0$^m$ &   0.52 &  $<0.04$   		   	& 2.470     (7) & 	       43.44  \\
  244 &   586 &     6047 &   53.164421 &  -27.842169 &  0.57   & 0 &   0.23 &   0.13    		   	  & 0.580     (1) & 	       41.46  \\
  245 &    27 &     8273 &   53.165272 &  -27.814062 &  0.23   & 0 &   6.34 &   0.85    		   	 & 3.064     (1) & 	       44.63  \\
  246 &    26 &    11588 &   53.165394 &  -27.769762 &  0.45   & 0 &   3.00 &   0.94    		   	& 1.552     (7) & 	        43.6  \\
  247 &    25 &      540 &   53.170151 &  -27.929647 &  0.14   & 0 &   7.53 &   0.62    		   	& 0.625     (1) & 	       43.11  \\
  249 &   611a &    14708 &   53.173809 &  -27.724491 &  0.65   & 0 &   1.05 &  $<0.09$   		   	& 0.979$^v$	(1) &         42.95   \\
  250 &    23 &    14158 &   53.174454 &  -27.733299 &  0.27   & 0 &   3.16 &   1.72    		   	& 2.576     (6) & 	       44.12  \\
  251 &    24 &     4302 &   53.174389 &  -27.867353 &  0.23   & 0 &   4.37 &   1.83    		   	& 3.610     (1) & 	       44.59  \\
  254 &    91 &    10429 &   53.178478 &  -27.784033 &  0.13   & 0 &   1.84 &   0.70    		   	& 3.193     (1) & 	       44.11  \\
  255 &   256 &     8374 &   53.179325 &  -27.812527 &  0.20   & 0 &   2.73 &   0.10    		   	& 1.730     (7) & 	       43.76  \\
  256 &    22 &     7822 &   53.180149 &  -27.820604 &  0.05   & 0 &   7.36 &   3.97    		   	& 1.920     (1) & 	       44.19  \\
  258 &    94c &    11006 &   53.183357 &  -27.776384 &  0.70$^{\dag}$   & 0 &   0.70 &   0.60   	   	& 2.688     (1) & 	       43.51  \\
\hline
\end{tabular}
\end{table*}}

\footnotesize{

\begin{table*}
%\caption{IRAC counterparts to CDFS X-ray sources - positions}\label{cptable}
%\centering
\vspace{0.2cm}
\begin{minipage}{0.99\textwidth}
\begin{tabular}{rrrccrrrrrr} 
\hline\hline             
AID & XID & OID & RA & DEC & dist & FLAG & S$_{2-10}$ keV & S$_{0.5-2}$ keV & redshift & log(L$_{X}$) \\ 
    &     &     & J2000 & J2000 & arcsec & & 10$^{-15}$ cgs & 10$^{-15}$  cgs &  & erg s$^{-1}$ \\
\hline
  259 &   132* &     1203 &   53.183304 &  -27.915039 &  0.89   & 0 &   0.90 &   0.27                & 0.908	 (1) &                    42.53 \\
  260 &    98a &     4578 &   53.184448 &  -27.861420 &  0.19   & 0 &  $<0.45$ &   0.53   	     & 0.279	 (1) & 		    41.3	\\
  261 &    21 &     3323 &   53.184639 &  -27.880918 &  0.27   & 0 &   0.86 &   0.58    	     & 3.471	 (1) & 		   43.84	\\
  262 &   573 &     8741 &   53.185150 &  -27.805275 &  0.35   & 0 &   0.25 &   0.14    	     & 0.414	 (1) & 		   41.16	\\
  263 &    20a &     7260 &   53.185230 &  -27.827835 &  0.19   & 0 &   4.79 &   0.64   	     & 1.016	 (1) & 		   43.39	\\
  264 &    85 &     8543 &   53.185833 &  -27.809969 &  0.23   & 0 &   1.45 &   0.52    	     & 2.593	 (1) & 		   43.81	\\
  266 &   646 &     9856 &   53.187954 &  -27.789999 &  0.40   & 0 &  $<0.34$ &   0.10   	     & 0.438	 (1) & 		   41.02	\\
  268 &   147* &    11062 &   53.193100 &  -27.775555 &  0.11   & 0$^m$ &   5.06 &   0.21   	     & 1.220	(3) &  		   43.68	\\
  269 &   170 &     1633 &   53.193371 &  -27.903877 &  0.31   & 0 &   1.34 &   0.25    	     & 0.664	 (1) & 		   42.38	\\
  272 &   146 &     2532 &   53.196079 &  -27.892647 &  0.25   & 0$^m$ &   2.52 &   0.52   	     & 2.74    (1.25-3.6)    & 	   44.11	\\
  273 &   592 &     4564 &   53.196575 &  -27.863205 &  0.42   & 0 &  $<0.55$ &   0.10   	     & 1.114	(3) &  		   42.34	\\
  276 &   184 &     3047 &   53.200737 &  -27.882391 &  0.12   & 0 &   1.68 &   0.18    	     & 0.667	 (1) & 		   43.15	\\
  277 &    17a &     1059 &   53.204872 &  -27.917955 &  0.44   & 0 &   2.15 &   0.86   	     & 2.02   (1.4-2.75)     & 	   43.72	\\
  281 &   159 &     3320 &   53.209366 &  -27.881090 &  0.33   & 0 &   8.19 &   2.67    	      & 3.470	 (4) & 		   44.84	\\
  283 &   508 &    70435 &   53.215118 &  -27.870249 &  0.20   & 0$^{k,m}$ &   1.09 &   0.15         & $>1.50$        & 		   $>43.15$ \\
  286 &    15 &     5062 &   53.220356 &  -27.855511 &  0.27   & 0 &   6.41 &   3.15    	      & 1.227	  (1) & 		   43.68 \\
  292 &   553a &     2750 &   53.236080 &  -27.887917 &  0.75   & 0 &  $<0.78$ &   0.25   	     & 0.366	 (1) & 		   41.23	 \\
  300 &   240 &     4751 &   53.245850 &  -27.861208 &  1.03   & 0$^m$ &  $<0.95$ &   0.27   	     & 0.16    (0.0-2.0)    &  	   40.47	 \\
\hline																		 
   70 &   537 &    70340 &   53.040970 &  -27.837717 &  0.52   & 1 &  $<0.48$ &   0.11               	  & $>5.00$	   &		    $>43.5$ \\
   81 &   236 &     6685 &   53.047676 &  -27.834925 &  0.39   & 1 &   0.46 &   0.24    	       & 1.920    (4) & 		   42.26 \\
   89 &   257 &    70286 &   53.055836 &  -27.815449 &  0.48   & 1 &   1.28 &   0.08    	      & 2.27  (1.0-4.0)      & 	   43.68	 \\
  104 &   587 &    70357 &   53.063774 &  -27.843531 &  1.08   & 1 &   0.31 &   0.13    	       & 2.28  (1.6-3.8)      & 	      43 \\
  128$^a$ &   254 &    70139 &   53.082344 &  -27.755135 &  0.61   & 1 &   1.60 &   0.05   	      & 2.28  (1.4-4.2)      & 	   43.82	 \\
  135 &    96 &    80001 &  53.086806 & -27.872953 &  0.21   & 1 &   0.90 &   0.33   		      & -1.00	     & 		      -1	 \\
  140 &   902 &    70380 &   53.091476 &  -27.853201 &  0.75   & 1 &  $<0.44$ &   0.08   	       & $>1.30$	&    		   $>42.08$ \\
  144 &   570 &    70251 &   53.093704 &  -27.801287 &  0.32   & 1 &  $<0.35$ &   0.17   	       &  2.84 (0.4-4.4)&		   43.15 \\
  232 &   557 &    70091 &   53.158222 &  -27.733713 &  0.60   & 1$^r$ &  $<0.43$ &   0.09   	      & $>3.80$        & 		   $>43.17$ \\
  \hline
   74 &    ... &    11122 &   53.044952 &  -27.774378 &  0.21   & 2 &   $<0.44$ &   0.10   		       & 1.607  	(2) &     42.34	  \\	  %wolf
   85 &    ... &    16795 &   53.052723 &  -27.696955 &  0.18   & 2 &   $<0.83$ &   0.08   		      & 1.78   (1.4-2.3)     &    42.71   \\
   97 &    ... &    13053 &   53.060192 &  -27.749054 &  0.52   & 2 &   $<0.49$ &   0.04   		      & 0.738	       (2) & 	    41.6  \\	  	  %wolf
  107 &    ... &    11065 &   53.065842 &  -27.775141 &  0.63   & 2 &   $<0.37$ &  $<0.06$   		      & 1.023	       (2) & 	   42.34  \\	  	  %wolf
  119$^a$ &    ... &     8977 &   53.073456 &  -27.803286 &  1.11   & 2 &    0.27 &  $<0.08$    	      & 0.735	    (3)  & 	   41.83  \\
  132 &    ... &    11695 &   53.084652 &  -27.765228 &  0.65   & 2 &   $<0.36$ &   0.14$^{f}$ 		      & 0.231	       (2) & 	   41.74  \\	  	     %wolf
  138 &    ... &    70198 &   53.090870 &  -27.782350 &  0.94   & 2 &    0.29 &  $<0.03$   		       & $>1.80$	& 	   $>42.85$  \\
  142 &    ... &    70307 &   53.092407 &  -27.826687 &  0.69   & 2 &   $<0.43$ &   0.06   		       & $>1.75$	& 	   $>42.39$  \\
  148 &    ... &    13552 &   53.094093 &  -27.740505 &  0.64   & 2 &   $<0.49$ &  0.19$^{f}$  		      & 0.738	       (2) & 	   42.25  \\	  	    %wolf
  165 &    ... &    70449 &   53.105164 &  -27.875032 &  0.81   & 2 &   $<0.37$ &   0.06   			 & 2.92  (1.7-4.7)	& 42.81   \\
  178 &    ... &     2768 &   53.112713 &  -27.888447 &  0.27   & 2 &   $<0.33$ &   0.08   		      & 1.103	  (3) & 	   41.84  \\
  185 &    ... &     4402 &   53.122162 &  -27.865400 &  0.23   & 2 &   $<0.47$ &   0.07   		      & 0.511	(7) &  	   41.18	  \\
  190 &    ... &     3290 &   53.124699 &  -27.881210 &  0.31   & 2 &   $<0.45$ &   0.06   		       & 1.017    (3) & 	   41.86  \\	  		  %wolf
  199 &    ... &    13111 &   53.133503 &  -27.747704 &  0.29   & 2 &   $<0.44$ &   0.17$^{f}$ 		      & 0.895	       (2) & 	    41.8  \\	  		    %wolf
  203 &    ... &    17284 &   53.137443 &  -27.688057 &  0.59   & 2 &   $<0.96$ &   0.05   		      & 1.071	  (4) & 	   42.32  \\			  %wolf
  215$^{p}$ &    ... &     7453 &   53.146000 &  -27.825754 &  0.92   & 2 &   $<0.19$ &   0.05   		       & 2.22	 (1.7-2.6)    &   42.36   \\
  217 &    ... &    70437 &   53.146461 &  -27.870947 &  0.45   & 2$^r$ &   $<0.45$ &   0.18$^{f}$   		 & $>4.42$	  & 	   $>43.49$  \\
  225$^{p}$ &  ...   &     7443 &   53.151447 &  -27.825872 &  0.28   & 2$^s$ &   $<0.34$ &   0.06    	       & 0.70	     & 	   41.34	  \\
  239$^{p}$ &   ... &     3743 &   53.161980 &  -27.875204 &  1.57    & 2 &   $<0.34$ &   0.04   			& 2.01  (1.0-2.6)      &  42.45   \\
  248 &    ... &     9957 &   53.172524 &  -27.788107 &  1.25   & 2 &   $<0.33$ &   0.04   		      & 0.622	   (7) & 	   41.24  \\
  257 &    ... &     8053 &   53.181049 &  -27.817164 &  0.39   & 2 &   $<0.27$ &   0.05   		      & 1.603	  (4) & 	   42.29  \\
  265 &    ... &     1263 &   53.187576 &  -27.910996 &  1.02   & 2 &   $<0.68$ &   0.16   		      & 0.458	   (7) & 	   41.26  \\
  267 &    ... &      564 &   53.190334 &  -27.926205 &  0.46   & 2 &    0.24 &   0.12   		      & 0.102	 (3) & 	   39.81	  \\
\hline																		  
  183$^{p}$ &    ... &    70241 &   53.120380 &  -27.798939 &  0.73$^{\dag}$  &  2$^{aa}$ &  $<0.23$ &   0.04    & 2.94   (1.3-3.60)     &      42.62   \\
      &        &    9269 & 53.119881    & -27.798738 & 1.04 & 2 &  $<0.23$ &   0.04  & ..  & 	..						  \\
  186 &   264 &    70403 &   53.123787 &  -27.862678 &  0.31$^{\dag}$  &  1 &   1.77 &   0.06   & 2.47  (1.4-5.5)      &                  43.95   \\
      &       &     4606 & 53.123939 & -27.863092 & 1.54 & 0 &   1.77 &   0.06   & .. & ..							  \\
\hline
  282 &    ... &    ... &  ... & ... & ...$^{\ddag}$ &   $-$ &   1.50 &  $<0.12$   & -1.00   & ..     \\ 
\end{tabular}
\vspace{0.5cm}
\par
Notes: \\
AID = X-ray identifier from A03; XID = X-ray identifier from S04, OID = MUSIC 
identifier from Santini et al. (2009) \\
$^a$: X-ray positions corrected from that reported in A03 (by visual
inspection of the event file); \\ 
$\dag$ Tentative X-ray to IRAC association; in this case, also a ``secondary''
candidate counterpart is reported. \\ 
$\ddag$ optical/IRAC counterpart not associated. \\
Legenda for the FLAG column: 
$0$ - counterpart published in S04; $1$ - counterpart different
from the one published in S04; $2$ - new published counterparts (sources
detected only in A03). \\
Notes on single sources:  $^k$: source published in
Koekemoer et al.(2004); $^m$: source published in Mainieri et al.(2005); $^r$:
source published in Rodighiero et al.(2007); $^{aa}$: source published by
Alonso-Herrero et al. (2006); $^s$: this source is listed as
counterpart ``b'' of XID 557 in S04 (577b).\\
$^f$: flux in the 0.5-8 keV band.   \\
$^p$: possible spurious source (see discussion in Sect. 2.1)
The redshift column report the photometric or spectroscopic redshift. When a spectroscopic redshift is present, the relative
source catalog is reported, as follow: (1) = CXO-CDFS (S04); (2) K20 (Mignoli et al. 2005); 
(3) GOODS-FORS2 (various releases); (4) GOODS-VIMOS (Popesso et al. 2008); (5)
COMBO-17 (Wolf et al. 2003); (6) GMASS; (7) others: VVDS, Daddi, TrzK, LCIRS ast0612152
Otherwise, the redshift is from SED fitting. 
(CCC): Photometric redshift soultion preferred to the low-quality spectroscopic redshift available \\
$^{v}$: this object has a different redshift solution from the VIMOS
observation, z=0.798 (Popesso et al. 2008)
\end{minipage}
\end{table*}}

\begin{table*}
\caption{Multiwavelength photometry for counterparts to CDFS X-ray sources}
{\scriptsize
\begin{tabular}{llcccccccccccc}
\hline\hline
AID & OID & b & v & i & z & j & h & k & S(3.6) & S(4.5) & S(5.8) & S(8.0) & S(24) \\
\hline
51 &    13272 &  26.51 &  26.27 &  25.53 &  25.17 &  23.99 &  23.24 &  23.06 &  22.33 &  22.15 &  22.16 &  22.44 & $>20.93$       	\\
53 &    12222 &  25.53 &  23.12 &  21.63 &  21.13 &  20.61 &  20.07 &  19.67 &  19.75 &  20.16 &  20.44 &  21.64 & $>22.46$       	\\
67 &    70103 &  28.06 &  28.03 &  26.72 &  26.78 & $>26.07$ &  23.76 &  23.67 &  22.60 &  22.43 &  22.50 &  22.83 &  22.56       	\\
71 &    70051 &  26.28 &  24.47 &  23.19 &  22.51 &  22.38 &  21.38 &  20.30 &  19.31 &  19.57 &  19.23 &  18.90 &  17.38        	\\
73 &    15717 &  21.81 &  19.97 &  19.06 &  18.68 &  18.17 &    ... &  17.58 &  18.04 &  18.30 &  18.68 &  19.10 &  18.86        	\\
75 &    14349 &  23.14 &  22.60 &  21.62 &  20.97 &  20.66 &  20.10 &  19.76 &  19.09 &  19.43 &  19.75 &  20.00 &  18.55        	\\
76 &    13849 &  23.70 &  22.96 &  22.38 &  22.39 &  21.66 &  21.41 &  20.93 &  20.02 &  19.65 &  19.33 &  19.11 &  17.91        	\\
77 &     8168 &  27.75 &  26.92 &  26.01 &  25.06 &  24.04 &  23.07 &  22.08 &  21.56 &  21.42 &  20.97 &  20.67 &  18.95        	\\
80 &    10619 &  24.25 &  22.74 &  21.66 &  21.30 &  20.69 &  20.49 &  20.24 &  20.32 &  20.64 &  20.78 &  21.25 &  20.16        	\\
82 &    11180 &  26.89 &  26.43 &  25.57 &  25.08 &  23.55 &  23.19 &  23.11 &  21.96 &  21.28 &  21.01 &  20.47 &  19.00        	\\
83 &    12281 &  24.59 &  22.47 &  21.15 &  20.74 &  20.31 &  19.87 &  19.48 &  19.61 &  20.05 &  20.27 &  20.94 &  20.20        	\\
84 &    11285 &  24.81 &  23.67 &  22.60 &  21.92 &  21.14 &  20.80 &  20.31 &  19.69 &  19.85 &  20.23 &  20.30 &  18.03        	\\
88 &    15626 &  21.83 &  20.92 &  20.06 &  19.84 &  19.25 &    ... &  18.63 &  18.60 &  18.85 &  18.74 &  18.81 &  16.82        	\\
90 &    12434 &  25.23 &  24.39 &  23.32 &  22.48 &  21.33 &  20.64 &  20.04 &  19.54 &  19.49 &  19.90 &  20.58 &  19.48        	\\
91 &    15382 &  23.41 &  22.36 &  21.27 &  20.83 &  20.04 &    ... &  19.22 &  19.15 &  19.50 &  19.48 &  19.39 &  17.74        	\\
93 &    70323 & $>28.37$ &  27.06 &  26.96 &  26.02 &  23.52 &  21.71 &  21.29 &  20.19 &  20.05 &  19.76 &  19.84 &  21.44      	\\
94 &     5434 &  23.95 &  23.10 &  22.05 &  21.57 &  21.08 &  20.65 &  20.28 &  20.14 &  20.87 &  20.36 &  20.34 &  18.93        	\\
95 &    16040 &  24.79 &  25.04 &  24.60 &  24.50 &  24.22 &    ... &  23.48 &  22.42 &  21.82 &  21.59 &  21.65 &  19.32        	\\
96 &     5281 &  24.34 &  23.94 &  23.38 &  22.86 &  21.68 &  21.26 &  21.05 &  22.95 &  20.29 &  20.64 &  20.90 &  19.25        	\\
        99 &    14104 &  27.04 &  26.66 &  25.61 &  25.13 &  24.02 &  23.10 &  22.75 &  21.68 &  21.51 &  21.60 &  21.83 &  20.08       \\
       100 &     5440 & $>29.21$ &  27.35 &  25.96 &  24.91 &  23.19 &  22.75 &  22.03 &  21.09 &  21.64 &  21.27 &  21.42 &  20.54     \\
       102 &    16094 &  24.88 &  23.67 &  22.95 &  22.15 &  20.30 &    ... &  19.66 &  19.08 &  18.95 &  19.05 &  19.57 &  19.87       \\
       103 &     4878 &  23.80 &  22.28 &  20.93 &  20.47 &  19.89 &    ... &  19.14 &  19.14 &  20.21 &  19.60 &  20.11 &  19.69       \\
       106 &     2654 &  24.75 &  23.26 &  22.17 &  21.73 &  21.12 &    ... &  20.29 &  19.98 &  20.83 &  20.45 &  20.62 &  18.23       \\
       111 &     4384 &  26.42 &  25.94 &  25.74 &  25.47 &  24.66 &    ... &  23.92 &  23.16 &  24.92 &  22.79 &  22.91 &  21.58       \\
       112 &    10944 &  27.25 &  26.63 &  26.17 &  25.89 & $>26.74$ &  25.03 & $>26.08$ &  24.56 &  23.95 &  23.91 &  23.12 & $>21.14$ \\
       113 &    14774 &  24.86 &  22.96 &  22.02 &  21.74 &  21.27 &    ... &  20.91 &  21.24 &  21.55 &  21.80 &  22.36 & $>21.11$     \\
       114 &     6606 &  26.18 &  24.35 &  22.92 &  22.53 &  21.95 &  21.67 &  21.43 &  21.47 &  22.03 &  22.23 &  22.67 & $>21.12$     \\
       115 &     7393 &  22.04 &  20.30 &  19.47 &  19.13 &  18.64 &  18.22 &  18.08 &  18.49 &  18.76 &  19.13 &  18.87 &  18.09       \\
       116 &    17038 &  27.45 &  26.59 &  26.11 &  25.64 &  23.96 &    ... &  22.82 &  21.92 &  21.57 &  21.42 &  21.61 &  19.49       \\
       117 &    15207 &  23.11 &  21.76 &  20.86 &  20.58 &  20.01 &    ... &  19.36 &  19.31 &  19.41 &  19.35 &  19.09 &  16.05       \\
       118 &     3920 &  24.97 &  23.87 &  22.65 &  21.80 &  21.03 &    ... &  19.93 &  19.46 &  19.86 &  19.70 &  19.52 &  17.89       \\
       121 &     9792 &  24.14 &  21.61 &  19.95 &  19.35 &  18.55 &  18.08 &  17.74 &  17.60 &  18.18 &  18.37 &  19.05 &  20.10       \\
       122 &     3533 &  24.51 &  24.36 &  24.25 &  24.30 &  24.01 &    ... &  23.41 &  22.68 &  22.55 &  22.19 &  22.34 &  21.46       \\
       123 &     5551 &  24.92 &  24.32 &  23.35 &  22.60 &  21.18 &  20.75 &  20.39 &  19.80 &  19.87 &  20.13 &  20.69 &  19.37       \\
       124 &    13866 &  24.91 &  23.92 &  22.74 &  22.08 &  21.37 &  20.82 &  20.37 &  20.26 &  20.57 &  20.95 &  21.30 &  19.88       \\
       125 &     4835 &  27.75 &  25.68 &  25.20 &  25.08 &  24.31 &    ... &  22.50 &  21.75 &  21.74 &  21.57 &  21.25 &  20.01       \\
       126 &     9279 &  26.46 &  24.92 &  23.77 &  23.45 &  22.83 &  22.71 &  22.49 &  22.23 &  22.33 &  22.03 &  21.63 &  20.08       \\
       129 &    17067 &  21.04 &  19.60 &  18.97 &  18.73 &  18.34 &    ... &  17.95 &  18.28 &  18.03 &  17.84 &  17.18 &  15.15       \\
       130 &    70052 &  27.65 &  28.44 &  26.86 &  26.25 &  24.86 &  23.57 &  22.97 &  23.05 &  23.01 &  22.51 &  23.18 & $>21.01$     \\
       131 &    13250 &  25.95 &  25.32 &  24.68 &  24.30 &  23.40 &  22.81 &  22.34 &  21.35 &  21.03 &  20.73 &  20.27 &  17.63       \\
133 &    70044 & $>28.56$ & $>28.86$ & $>28.07$ &$>27.96$ &$>26.00$ &$>25.06$ &$>25.25$ &  23.49 &  23.05&22.56 & 22.15 & $>21.01$      \\
       134 &     9723 &  24.95 &  23.31 &  22.11 &  21.62 &  21.02 &  20.24 &  19.88 &  19.73 &  20.15 &  20.13 &  20.38 &  19.15       \\
       136 &      489 &  25.55 &  25.01 &  24.67 &  24.45 &  23.81 &    ... &  23.37 &  22.75 &  23.13 &  22.16 &  22.68 &  20.51       \\
       137 &    15948 &  23.43 &  22.56 &  21.53 &  21.16 &  20.50 &  19.97 &  19.86 &  19.81 &  20.28 &  20.32 &  20.76 &  18.71       \\
       139 &    10384 &  23.77 &  22.27 &  20.92 &  20.46 &  19.76 &  19.25 &  19.02 &  18.94 &  19.43 &  19.56 &  20.11 &  18.62       \\
       141 &    70249 &  27.27 &  28.91 &  28.08 & $>27.96$ &  24.60 &  23.59 &  22.76 &  21.30 &  21.00 &  20.47 &  20.47 &  18.21     \\
       143 &    12146 &  27.70 &  27.70 &  26.87 &  26.57 &  25.85 &  25.15 &  24.75 &  23.78 &  23.55 &  23.36 &  23.74 & $>21.50$     \\
       145 &    11723 &  25.18 &  25.09 &  24.57 &  24.32 &  23.34 &  22.91 &  22.33 &  21.11 &  20.54 &  20.07 &  19.61 &  18.64       \\
       146 &     7031 &  26.18 &  24.12 &  22.62 &  22.03 &  21.09 &  20.48 &  19.92 &  19.72 &  20.26 &  20.30 &  20.74 &  19.02       \\
       149 &     3781 &  21.21 &  20.41 &  19.99 &  19.83 &  19.67 &    ... &  19.46 &  20.07 &  20.89 &  20.93 &  20.14 &  18.77       \\
       151 &    11854 &  24.65 &  23.95 &  23.13 &  22.33 &  21.76 &  21.25 &  20.93 &  20.34 &  20.36 &  20.78 &  20.76 &  18.81       \\
       152 &     8029 &  20.80 &  20.10 &  19.86 &  19.80 &  19.62 &  19.68 &  19.57 &  20.15 &  20.39 &  20.80 &  19.85 &  19.03       \\
       154 &    15419 &  25.14 &  24.68 &  24.47 &  24.17 &  23.25 &  22.27 &  21.86 &  21.28 &  21.07 &  20.96 &  21.28 &  20.05       \\
       155 &    17077 &  23.34 &  21.66 &  20.54 &  20.11 &  19.58 &  19.08 &  18.80 &  19.06 &  19.32 &  19.48 &  19.89 &  19.31       \\
       157 &     1780 &  24.77 &  22.80 &  21.52 &  21.10 &  20.39 &    ... &  19.73 &  19.67 &  20.70 &  20.33 &  20.91 &  19.84       \\
       158 &    10250 &  22.11 &  21.06 &  20.59 &  20.32 &  20.10 &  19.48 &  19.58 &  19.95 &  19.95 &  20.33 &  19.18 &  18.13       \\
       159 &      371 &  25.60 &  24.95 &  23.90 &  23.20 &  22.30 &    ... &  21.35 &  20.74 &  21.03 &  20.97 &  21.40 &  20.12       \\
       160 &     6640 &  25.80 &  25.43 &  24.41 &  23.97 &  23.34 &  23.02 &  22.80 &  22.34 &  22.61 &  22.52 &  22.36 & $>22.28$     \\
       162 &     5832 &  24.79 &  24.24 &  23.25 &  22.67 &  22.27 &  22.13 &  21.89 &  21.48 &  21.85 &  22.02 &  22.08 &  20.24       \\
       163 &    16224 &  24.79 &  23.62 &  22.64 &  22.21 &  21.59 &  20.96 &  21.05 &  20.33 &  19.70 &  19.38 &  19.01 &  17.53       \\
       164 &     1192 &  24.76 &  23.53 &  22.17 &  21.21 &  20.48 &    ... &  19.64 &  18.98 &  19.63 &  19.71 &  20.14 &  19.04       \\
       166 &    15260 &  26.62 &  26.13 &  25.44 &  25.02 &  23.61 &  22.29 &  21.89 &  20.79 &  20.11 &  19.25 &  18.37 &  17.12       \\
       167 &     7194 &  22.54 &  21.10 &  20.04 &  19.66 &  19.13 &  18.75 &  18.54 &  18.55 &  19.03 &  19.24 &  19.90 &  19.45       \\
\hline
\end{tabular}}
\end{table*}

\begin{table*}
{\scriptsize 
\begin{tabular}{llcccccccccccc}
\hline\hline
AID & OID & b & v & i & z & j & h & k & S(3.6) & S(4.5) & S(5.8) & S(8.0) & S(24) \\ 
\hline
       170 &    70385 & $>28.60$ &  27.77 &  28.01 &  26.49 &  25.73 & $>25.26$ &  22.65 &  22.56 &  22.37 &  22.08 &  22.16 & $>21.39$ \\
       171 &      978 &  24.19 &  23.73 &  23.00 &  22.55 &  22.09 &    ... &  21.77 &  21.25 &  21.63 &  21.51 &  21.75 &  19.48       \\
       172 &    12747 &  27.75 &  26.61 &  26.07 &  25.95 &  25.05 &  23.99 &  23.42 &  23.01 &  22.78 &  22.54 &  21.68 &  21.72        \\
       175 &    70168 & $>28.40$ &  28.20 &  26.54 &  25.97 & $>25.81$ & $>25.03$ &  24.03 &  22.92 &  23.00 &  22.04 &  21.77 &  20.63 \\
       176 &    16792 &  24.39 &  23.08 &  21.87 &  21.44 &  20.81 &  20.23 &  19.94 &  19.49 &  19.59 &  19.40 &  19.20 &  17.43       \\
       177 &    17452 &  19.40 &  19.31 &  19.13 &  19.00 &  18.77 &  18.68 &  18.41 &  17.51 &  17.21 &  17.03 &  16.89 &  16.56       \\
       179 &    16831 &  24.05 &  22.81 &  21.62 &  21.31 &  20.91 &  20.33 &  20.10 &  19.65 &  19.56 &  19.26 &  18.96 &  17.50       \\
       180 &    10464 &  25.14 &  24.78 &  24.40 &  24.18 &  23.33 &  22.73 &  22.69 &  21.68 &  21.55 &  21.30 &  21.59 &  19.04        \\
       181 &    11248 &  25.13 &  23.17 &  21.70 &  21.15 &  20.60 &  19.72 &  19.70 &  19.63 &  20.11 &  20.30 &  20.92 & $>21.08$     \\
       182 &    13977 &  20.50 &  18.85 &  18.18 &  17.87 &  17.52 &  17.34 &  17.19 &  17.79 &  18.06 &  18.67 &  18.98 &  19.81       \\
       187 &     2637 &  29.66 &  27.59 &  26.94 &  26.47 &  25.32 &    ... &  22.99 &  21.73 &  21.62 &  21.26 &  21.55 &  20.22        \\
       188 &     5390 &  27.52 &  25.05 &  24.53 &  24.62 &  24.50 &  23.44 &  22.49 &  21.91 &  21.81 &  21.62 &  21.48 &  19.20       \\
       189 &    13855 &  17.57 &  16.86 &  16.45 &  16.30 &  16.01 &  15.87 &  15.92 &  16.14 &  16.41 &  15.97 &  13.86 &  14.06       \\
       191 &    12354 &  21.15 &  21.19 &  21.12 &  21.07 &  20.84 &  20.60 &  20.20 &  19.39 &  19.08 &  18.94 &  18.62 &  17.33       \\
       192 &    13853 &  18.16 &  17.39 &  16.95 &  16.76 &  16.45 &  16.28 &  16.32 &  16.90 &  16.96 &  16.68 &  14.73 &  14.64       \\
       193 &    12489 &  23.45 &  22.87 &  22.23 &  21.74 &  21.11 &  20.82 &  20.21 &  19.78 &  19.82 &  19.89 &  19.73 &  17.49       \\
       195 &    12880 &  22.61 &  22.10 &  21.72 &  21.51 &  21.27 &  21.22 &  20.93 &  20.64 &  20.68 &  20.52 &  20.56 &  18.60       \\
       196 &     9791 &  22.57 &  21.90 &  21.24 &  21.08 &  20.68 &  20.14 &  20.06 &  20.22 &  20.62 &  20.50 &  20.79 &  18.60       \\
       197 &    11272 &  24.37 &  23.68 &  23.21 &  22.92 &  22.06 &  21.07 &  20.81 &  20.33 &  20.01 &  19.61 &  19.27 &  17.66       \\
       198 &     8217 &  25.90 &  25.43 &  25.21 &  24.87 &  23.70 &  23.15 &  22.86 &  21.30 &  21.00 &  20.85 &  21.21 &  19.50       \\
       200 &    16661 &  24.71 &  24.17 &  23.53 &  23.10 &  22.43 &  21.85 &  21.50 &  20.68 &  20.33 &  20.07 &  20.31 &  17.72       \\
       201 &    70195 & $>28.71$ &  27.02 &  26.89 &  26.14 & $>25.72$ &  23.61 &  23.34 &  22.36 &  22.39 &  21.71 &  21.33 &  20.16    \\
       202 &    70404 &  24.53 &  23.79 &  23.00 &  22.43 &  21.45 &  20.87 &  20.44 &  19.47 &  19.68 &  19.69 &  19.96 &  20.07       \\
       204 &    11970 &  24.27 &  23.06 &  21.89 &  21.26 &  20.36 &  19.88 &  19.22 &  19.07 &  19.26 &  19.48 &  19.57 &  17.54       \\
       205 &    70429 & $>28.63$ & $>28.82$ & $>28.04$ & $>27.90$ & $>25.96$ &  25.41 &  23.86 &  23.27 &  22.98 &  22.49 &  22.48 & $>21.01$ \\ 
       206 &    15927 & $>29.15$ &  27.19 &  25.75 &  24.63 &  23.45 &  22.60 &  22.41 &  22.10 &  21.52 &  21.18 &  20.60 &  20.09     \\
       207 &    14880 &  18.78 &  17.48 &  16.90 &  16.62 &  16.27 &  15.94 &  16.09 &  17.56 &  17.12 &  17.60 &  17.51 &  17.94       \\
       208 &    12616 &  26.53 &  25.76 &  25.39 &  25.09 &  24.16 &  23.69 &  23.10 &  22.78 &  22.45 &  22.39 &  22.37 &  21.36       \\
       209 &    70289 &  27.46 &  27.06 &  26.68 &  26.22 &  25.43 &  23.60 &  22.91 &  21.70 &  21.43 &  21.02 &  21.08 &  19.36        \\
       210 &     7316 &  27.33 &  25.14 &  23.60 &  22.89 &  22.26 &  21.74 &  21.34 &  21.18 &  21.64 &  21.84 &  22.39 & $>21.13$     \\
       212 &    14262 &  24.61 &  22.79 &  21.49 &  21.07 &  20.51 &  20.23 &  19.95 &  20.48 &  20.38 &  20.60 &  21.20 &  20.29       \\
       214 &      921 &  23.76 &  23.15 &  22.21 &  21.79 &  20.92 &    ... &  20.23 &  20.14 &  20.36 &  20.20 &  20.07 &  18.18       \\
       216 &    13946 &  25.11 &  24.49 &  23.86 &  23.54 &  22.43 &  21.99 &  21.44 &  21.83 &  20.58 &  20.53 &  21.02 &  18.50       \\
       218 &     2818 &  28.54 &  28.18 &  27.03 &  26.06 &  24.61 &    ... &  23.45 &  22.42 &  22.25 &  21.84 &  21.67 &  20.02        \\
       219 &     7814 &  25.68 &  25.39 &  25.00 &  24.88 &  23.81 &  22.36 &  22.01 &  21.16 &  20.75 &  20.51 &  19.77 &  17.00       \\
       221 &     8231 &  22.42 &  22.23 &  22.00 &  21.64 &  21.44 &  21.09 &  20.95 &  20.48 &  20.51 &  20.88 &  20.91 &  19.17       \\
       222 &     4993 &  25.97 &  25.36 &  24.43 &  23.91 &  22.99 &    ... &  21.73 &  20.81 &  20.70 &  20.86 &  21.14 &  18.99       \\
       223 &     5988 &  29.38 &  28.02 &  26.81 &  25.84 &  23.68 &  22.86 &  22.39 &  21.48 &  21.57 &  21.49 &  21.97 &  21.16       \\
       224 &     7347 &  22.77 &  21.48 &  20.62 &  20.24 &  19.67 &  19.05 &  18.92 &  18.86 &  19.11 &  19.05 &  18.74 &  16.81       \\
       226 &    11049 &  24.99 &  23.88 &  22.69 &  22.13 &  21.51 &  21.20 &  20.77 &  20.52 &  20.83 &  21.09 &  21.12 &  19.12       \\
       227 &    13964 &  24.43 &  23.18 &  22.05 &  21.64 &  21.06 &  20.74 &  20.41 &  20.96 &  20.38 &  20.38 &  20.48 &  19.08       \\
       230 &     4119 &  25.15 &  24.72 &  24.08 &  23.52 &  21.68 &    ... &  20.59 &  19.56 &  19.17 &  18.57 &  17.99 &  16.44       \\
       231 &     3010 &  26.86 &  25.91 &  25.96 &  25.76 &  24.52 &    ... &  23.45 &  22.45 &  22.34 &  22.15 &  22.13 &  20.36       \\
       233 &    11237 & $>30.06$ &  28.70 &  27.60 &  26.48 &  24.91 &  23.77 &  22.98 &  22.04 &  21.79 &  21.30 &  20.62 &  19.35     \\
       235 &      461 &  28.37 &  27.97 &  26.71 &  25.91 &  23.84 &    ... &  22.22 &  21.21 &  20.98 &  21.06 &  21.34 &  18.65       \\
       236 &     9696 &  21.97 &  21.53 &  21.18 &  20.93 &  20.72 &  20.31 &  20.07 &  20.19 &  20.14 &  20.16 &  19.32 &  16.81       \\
       237 &     5099 &  27.81 &  26.33 &  26.14 &  26.11 &  25.10 &    ... &  23.55 &  22.88 &  22.82 &  22.46 &  21.99 &  19.62       \\
       238 &     6757 &  21.99 &  20.69 &  19.96 &  19.64 &  18.90 &  18.35 &  17.93 &  18.62 &  18.64 &  19.00 &  17.23 &  17.11       \\
       241 &    70116 &  27.05 &  26.68 &  26.44 &  26.06 & $>25.89$ &  23.40 &  23.23 &  22.66 &  22.49 &  22.09 &  21.89 & $>21.85$   \\
       242 &    11701 &  21.26 &  21.02 &  20.96 &  20.99 &  21.10 &  20.55 &  20.59 &  19.42 &  19.27 &  19.18 &  18.97 &  17.97       \\
       243 &     8625 & $>29.64$ &  27.58 &  26.36 &  26.21 &  25.18 &  23.37 &  22.87 &  22.33 &  22.20 &  22.07 &  22.17 & $>22.61$   \\
       244 &     6047 &  25.21 &  22.96 &  21.65 &  21.21 &  20.62 &  20.13 &  19.91 &  19.99 &  20.52 &  20.64 &  21.35 & $>21.48$     \\
       245 &     8273 &  28.93 &  25.68 &  24.70 &  24.51 &  23.77 &  22.40 &  21.67 &  21.05 &  20.84 &  20.61 &  20.31 &  18.69       \\
       246 &    11588 &  26.02 &  25.83 &  25.30 &  24.93 &  24.33 &  23.24 &  22.91 &  20.86 &  20.79 &  20.35 &  20.51 &  18.28       \\
       247 &      540 &  27.89 &  26.15 &  25.48 &  25.37 &  24.57 &    ... &  21.82 &  21.29 &  20.81 &  20.34 &  19.19 &  17.04       \\
       249 &    14708 &  25.31 &  24.40 &  23.24 &  22.54 &  21.67 &    ... &  20.49 &  19.95 &  20.06 &  20.30 &  20.32 &  18.31       \\
       250 &    14158 &  26.64 &  25.32 &  25.01 &  24.60 &  24.06 &  23.71 &  23.46 &  22.57 &  22.29 &  21.94 &  21.89 &  22.16       \\
       251 &     4302 &  24.35 &  22.89 &  22.41 &  22.43 &  21.74 &    ... &  21.40 &  21.23 &  21.44 &  21.17 &  20.82 &  18.95       \\
       254 &    10429 &  26.47 &  25.05 &  25.18 &  25.18 &  24.49 &  24.08 &  23.74 &  23.21 &  23.15 &  22.85 &  22.79 &  21.04       \\
       255 &     8374 &  26.07 &  25.80 &  25.08 &  24.50 &  22.88 &  22.34 &  22.01 &  21.32 &  21.23 &  21.29 &  21.64 &  20.35       \\
       256 &     7822 &  22.42 &  22.91 &  22.55 &  22.50 &  22.09 &    ... &  21.47 &  20.75 &  20.53 &  20.28 &  20.27 &  18.61       \\
       258 &    11006 &  25.20 &  24.66 &  24.52 &  24.44 &  24.33 &  23.85 &  23.59 &  21.93 &  21.70 &  21.17 &  20.78 &  18.78       \\
\hline
\end{tabular}}
\end{table*}

\vspace{0.2cm}
\begin{table*}
\begin{minipage}{0.99\textwidth}
{\scriptsize 
\begin{tabular}{llcccccccccccc}
\hline\hline
AID & OID & b & v & i & z & j & h & k & S(3.6) & S(4.5) & S(5.8) & S(8.0) & S(24) \\ 
\hline
       259 &     1203 &  25.68 &  25.30 &  24.53 &  24.33 &  23.93 &    ... &  23.52 &  22.58 &  22.59 &  22.36 &  22.13 &  22.01       \\
       260 &     4578 &  21.64 &  20.19 &  19.47 &  19.15 &  18.48 &    ... &  17.92 &  18.21 &  18.09 &  17.76 &  16.80 &  14.35       \\
       261 &     3323 &  25.01 &  23.99 &  23.87 &  23.98 &  23.72 &    ... &  23.37 &  22.53 &  22.39 &  21.96 &  21.64 &  20.95       \\
       262 &     8741 &  22.33 &  21.19 &  20.50 &  20.15 &  19.83 &  19.39 &  19.15 &  19.57 &  19.68 &  20.00 &  19.23 &  18.14       \\
       263 &     7260 &  25.49 &  24.38 &  23.02 &  22.32 &  21.62 &    ... &  21.06 &  20.71 &  21.05 &  21.18 &  21.38 &  19.78       \\
       264 &     8543 &  25.58 &  25.12 &  24.68 &  24.61 &  24.32 &  23.57 &  23.20 &  22.08 &  21.82 &  21.49 &  21.43 &  19.72       \\
       266 &     9856 &  22.74 &  21.58 &  20.84 &  20.45 &  20.04 &  19.50 &  19.20 &  19.60 &  19.70 &  19.90 &  18.98 &  17.36       \\
       268 &    11062 &  26.37 &  25.73 &  24.66 &  23.97 &  22.96 &  22.44 &  22.08 &  21.38 &  21.44 &  21.51 &  20.88 &  19.36       \\
       269 &     1633 &  23.08 &  21.29 &  20.00 &  19.55 &  19.01 &    ... &  18.12 &  17.83 &  18.42 &  18.47 &  19.07 &  17.73       \\
       272 &     2532 &  28.83 &  27.70 &  26.38 &  26.00 &  24.48 &    ... &  22.67 &  22.16 &  22.17 &  21.99 &  22.02 &  22.06       \\
       273 &     4564 &  24.42 &  23.83 &  23.00 &  22.25 &  21.58 &    ... &  20.82 &  20.27 &  20.68 &  20.98 &  21.24 &  19.87       \\
       276 &     3047 &  23.60 &  21.62 &  20.27 &  19.86 &  19.29 &    ... &  18.47 &  18.28 &  18.89 &  18.99 &  19.53 &  18.38       \\
       277 &     1059 &  27.72 &  26.81 &  25.95 &  24.96 &  23.65 &    ... &  21.99 &  20.93 &  20.79 &  20.40 &  20.62 & $>21.53$     \\
       281 &     3320 &  26.99 &  25.63 &  25.20 &  25.29 &  24.51 &    ... &  23.03 &  23.48 &  22.27 &  22.61 &  20.44 &  18.50       \\
       283 &    70435 &  28.44 &  28.00 &  27.26 &  26.88 & $>25.97$ &  24.06 &  23.95 &  22.51 &  22.18 &  21.63 &  21.30 &  19.25     \\
       286 &     5062 &  23.64 &  22.86 &  22.26 &  21.80 &  21.30 &    ... &  20.39 &  19.70 &  19.88 &  19.82 &  19.78 &  18.66       \\
       292 &     2750 &  20.57 &  20.16 &  20.27 &  19.91 &  20.11 &    ... &  19.71 &  20.21 &  20.25 &  20.40 &  19.58 &  16.88       \\
       300 &     4751 &  28.88 &  28.29 &  27.19 &  26.89 &  26.33 &    ... &  25.44 &  22.66 & $>24.92$ &  21.71 &  21.04 & $>20.69$   \\
\hline                                                                                                                                  
        70 &    70340 &  26.63 &  28.16 &  27.41 & $>27.56$ &  25.52 & $>25.46$ & $>25.62$ &  23.40 &  23.06 &  22.01 &  21.34 &  19.01 \\
        81 &     6685 &  28.44 &  26.80 &  25.11 &  24.23 &  22.79 &  22.08 &  21.59 &  20.90 &  21.01 &  21.33 &  21.59 &  21.17       \\
        89 &    70286 &  27.64 &  27.07 &  26.70 &  26.93 &  24.82 & $>24.88$ &  23.49 &  23.89 &  24.07 & $>23.30$ & $>23.03$ &  20.88 \\
       104 &    70357 & $>28.83$ &  27.28 &  26.58 &  25.95 &  23.69 &  23.40 &  22.92 &  21.59 &  21.57 &  21.11 &  20.75 &  18.75     \\
       128 &    70139 &  27.58 &  26.77 &  27.09 &  26.32 &  24.65 &  24.37 &  23.76 &  22.51 &  22.50 &  22.12 &  21.96 &  19.90       \\
       135 &    80001 &  26.89 &  26.80 &  27.85 &  27.04 &  25.65 &  24.98 & >25.35 &  23.57 &  23.27 &  22.69 &  21.17 &  20.46    \\
       140 &    70380 &  28.33 & $>28.72$ & $>27.94$ & $>27.66$ &  25.01 &  24.19 &  23.00 &  21.68 &  21.30 &  20.89 &  21.16 & $>21.02$ \\
       144 &    70251 &  28.34 &  27.82 &  27.51 &  26.94 & $>25.88$ &  24.43 &  23.38 &  23.53 &  23.02 &  23.18 &  23.15 & $>21.02$   \\
       232 &    70091 & $>28.60$ &  27.92 &  28.01 &  27.65 &  25.32 &  24.55 &  23.53 &  22.31 &  22.09 &  21.67 &  21.21 &  19.68     \\
\hline
        74 &    11122 &  26.06 &  25.42 &  23.81 &  22.87 &  21.01 &  20.39 &  19.90 &  19.23 &  18.94 &  19.20 &  19.76 & $>21.09$     \\
        85 &    16795 &  27.05 &  26.49 &  25.70 &  25.20 &  23.70 &    ... &  22.83 &  21.97 &  21.76 &  21.95 &  21.96 &  21.30       \\
        97 &    13053 &  27.98 &  25.11 &  23.68 &  23.25 &  22.76 &  22.26 &  21.94 &  22.25 &  22.48 &  22.62 &  23.03 &  20.76       \\
       107 &    11065 &  23.94 &  23.62 &  23.09 &  22.89 &  22.43 &  22.46 &  22.17 &  22.37 &  21.33 &  22.91 &  22.48 &  19.38       \\
       119 &     8977 &  25.60 &  24.29 &  23.06 &  22.54 &  21.66 &  20.94 &  20.43 &  20.01 &  20.48 &  20.45 &  20.74 &  18.63       \\
       132 &    11695 &  21.44 &  20.65 &  20.31 &  20.21 &  20.06 &  19.88 &  19.72 &  20.41 &  20.60 &  20.95 &  20.23 &  19.76       \\
       138 &    70198 &  28.34 &  28.89 &  26.58 &  26.22 & $>25.40$ &  23.28 &  23.53 &  21.17 &  21.37 &  20.94 &  20.99 &  18.42     \\
       142 &    70307 & $>28.59$ &  27.24 &  27.89 &  26.43 &  24.19 &  23.66 &  22.75 &  21.75 &  21.80 &  21.12 &  20.71 &  19.50     \\
       148 &    13552 &  22.51 &  21.61 &  20.62 &  20.29 &  19.71 &  19.21 &  18.72 &  18.70 &  19.10 &  19.14 &  19.38 &  17.01       \\
       165 &    70449 & $>28.70$ & $>28.89$ & $>28.10$ &  27.70 &  24.82 &  23.57 &  22.70 &  21.30 &  20.88 &  20.37 &  20.22 &  18.54 \\
       178 &     2768 &  24.35 &  23.78 &  22.92 &  22.22 &  21.63 &    ... &  20.65 &  20.04 &  20.70 &  20.84 &  21.19 &  19.73       \\
       185 &     4402 &  22.90 &  22.54 &  22.22 &  22.21 &  22.19 &    ... &  21.97 &  22.33 &  22.72 &  23.00 &  22.97 &  21.21       \\
       190 &     3290 &  26.85 &  25.80 &  24.46 &  23.62 &  22.80 &    ... &  21.89 &  21.51 &  21.95 &  22.33 &  22.55 & $>21.49$     \\
       199 &    13111 &  25.25 &  23.73 &  22.26 &  21.59 &  20.88 &  20.59 &  20.14 &  20.01 &  20.32 &  20.67 &  21.23 &  21.01       \\
       203 &    17284 &  24.47 &  23.21 &  21.79 &  20.87 &  20.04 &  19.43 &  19.04 &  18.49 &  18.77 &  19.21 &  19.54 &  18.83       \\
       215 &     7453 &  25.12 &  24.89 &  24.73 &  24.76 &  24.24 &  23.78 &  23.90 &  23.27 &  23.22 &  23.45 & $>24.46$ &  21.32     \\
       217 &    70437 & $>28.69$ &  28.52 &  28.10 &  27.62 & $>26.02$ & $>25.35$ &  23.58 &  23.02 &  22.53 &  21.94 &  21.32 &  20.34 \\
       225 &     7443 &  27.33 &  26.16 &  25.05 &  24.33 &  22.82 &  22.12 &  21.63 &  21.07 &  20.87 &  21.17 &  22.11 & $>21.14$     \\
       239 &     3743 &  26.13 &  26.03 &  26.09 &  25.90 &  24.89 &    ... &  24.92 &  24.71 &  25.12 &  24.48 & $>23.57$ & $>21.51$   \\
       248 &     9957 &  24.93 &  22.61 &  21.16 &  20.71 &  20.08 &  19.58 &  19.27 &  19.41 &  19.94 &  20.17 &  20.91 &  21.47       \\
       257 &     8053 &  26.45 &  26.09 &  25.16 &  24.70 &  23.24 &    ... &  22.05 &  20.94 &  20.73 &  20.70 &  20.91 &  17.94       \\
       265 &     1263 &  20.85 &  19.89 &  19.29 &  18.99 &  18.69 &    ... &  18.09 &  18.21 &  18.37 &  18.57 &  17.77 &  16.42       \\
       267 &      564 &  20.80 &  20.48 &  20.26 &  20.36 &  20.32 &    ... &  20.22 &  20.83 &  21.17 &  21.27 &  20.65 &  17.79       \\
\hline	
       183 &    70241 & $>28.71$ &  27.11 &  26.37 &  25.79 & $>24.73$ &  23.25 &  22.76 &  21.85 &  21.94 &  21.57 &  21.40 &  20.07   \\
       186 &    70403 &  27.25 &  27.83 &  26.67 &  26.56 &  25.27 &  24.77 &  24.28 &  22.31 &  22.43 &  21.95 &  21.95 &  19.98       \\
\hline	
       282 &     ... &    ... &    ... &    ... &    ... &    ... &    ... &    ... &    ... &    ... &    ... &    ... &    ...        \\
\hline
\end{tabular}}
\vspace{0.5cm}
\par
Notes: \\
All the magnitudes in the AB system. \\
\end{minipage}
\end{table*}

\begin{acknowledgements}
The authors are grateful to P. N. Best for deriving the SDSS
OIII source fractions as a function of stellar mass with an optical  
luminosity cut that matches the present X--ray data.
MB and FF acknowledge useful discussions with Roberto Gilli, Paolo Tozzi, John 
Silverman, Cristian Vignali, G. Cesare Perola, Silvia Bonoli, Niel Brandt, 
Bin Luo, Hagai Netzer and Jochen Greiner. We thank the anonymous referee
for insightful and detailed comments which helped in the presentation of
the results. MB acknowledges support
from the  XMM--{\it Newton} DLP grant 50-)G-0502; 
FF and AC acknowledge support from ASI
contracts I/088/06/0 and I/016/07/0 and PRIN/MIUR 2006-02-5203.
 
\end{acknowledgements}

\end{document}